\documentclass[a4paper,11pt]{article}
\usepackage{jheppub}	
\usepackage[english]{babel}
\usepackage{makeidx}
\usepackage{amsfonts}
\usepackage{enumerate}
\usepackage{mathrsfs}
\usepackage{tensor}
\usepackage[autostyle]{csquotes}
\usepackage{subfig}
\usepackage{simpler-wick}

\newdimen\tableauside\tableauside=1.0ex
\newdimen\tableaurule\tableaurule=0.4pt
\newdimen\tableaustep
\def\phantomhrule#1{\hbox{\vbox to0pt{\hrule height\tableaurule
width#1\vss}}}
\def\phantomvrule#1{\vbox{\hbox to0pt{\vrule width\tableaurule
height#1\hss}}}
\def\sqr{\vbox{%
  \phantomhrule\tableaustep
\hbox{\phantomvrule\tableaustep\kern\tableaustep\phantomvrule\tableaustep}%
  \hbox{\vbox{\phantomhrule\tableauside}\kern-\tableaurule}}}
\def\squares#1{\hbox{\count0=#1\noindent\loop\sqr
  \advance\count0 by-1 \ifnum\count0>0\repeat}}
\def\tableau#1{\vcenter{\offinterlineskip
  \tableaustep=\tableauside\advance\tableaustep by-\tableaurule
  \kern\normallineskip\hbox
    {\kern\normallineskip\vbox
      {\gettableau#1 0 }%
     \kern\normallineskip\kern\tableaurule}%
  \kern\normallineskip\kern\tableaurule}}
\def\gettableau#1 {\ifnum#1=0\let\next=\null\else
  \squares{#1}\let\next=\gettableau\fi\next}

		\def\b{\beta}				
						
				\def\l{\lambda}		
									\def\r{\rho}	
		\def\t{\tau}						
\def\c{\chi}

\def\be{\begin{equation}}
\def\ee{\end{equation}}
\def\bea{\begin{eqnarray}}
\def\eea{\end{eqnarray}}

\title{  Entanglement Entropy of Interacting Scalar Theories on Fuzzy Spaces }

\author{A. Allouche,}
\author{D. Dou}

\affiliation{Dept of Physics, College of Exact Sciences, Hamma Lakhdar University, El Oued, Algeria.}
\emailAdd{allouche-amel,dou-djamel@univ-eloued.dz}

\abstract{We investigate the impact of self-interactions on the R\'{e}nyi and entanglement entropies of a scalar field on $(2+1)$-dimensional spacetimes, whose spatial sections are modeled by fuzzy spaces, specifically the fuzzy sphere and the fuzzy disc. We compute the first-order perturbative correction induced by a $\lambda\phi^4$ interaction using the Green's function approach. In contrast to the free theory, where the entanglement entropy is dominated by degrees of freedom near the entangling boundary and obeys an area law, we find that the interaction correction has an extensive bulk contribution, receiving significant contributions from degrees of freedom throughout the fuzzy space. For the fuzzy sphere, the correction exhibits strong infrared sensitivity associated with the zero mode. We isolate and resolve this zero-mode IR divergence by projecting out the zero mode, thereby obtaining a physically meaningful quantity. In the commutative continuum limit, the interaction correction has the same degree of UV divergence as the free entropy but does not obey a pure area law. Furthermore, we analyze the Moyal plane limit, where the interaction correction exhibits a distinct IR divergence. We discuss the physical origin of these extensive bulk features and examine their possible connection to the celebrated UV/IR mixing phenomenon in noncommutative quantum field theories.}


\preprint{PREPRINT}
 \clearpage

\begin{document}
\maketitle


\section{Introduction}
The entanglement entropy (EE)\footnote{Throughout this paper, we use EE and DF as abbreviations for entanglement entropy and degrees of freedom, respectively.} provides a natural and quantum statistical interpretation
for the area scaling law \cite{Bombelli:1986rw,Srednicki:1993im,Callan:1994py,Susskind:1994sm}. Although it is not necessary that the black hole entropy
	is the entanglement of vacuum fluctuations of quantum fields in nature, the latter
	must be present and any consistent quantum theory of spacetime must account
	for them. On the other hand, it is well known that the EE is
	divergent in ordinary quantum field theory due to the absence of an UV cutoff. The
	need of UV cutoff and the finiteness of black hole entropy are widely viewed as a
	direct manifestation for a discrete nature underlying spacetime at the Planck scale,
	and points out to a necessary reduction of the number of degrees of freedom on
	the horizon \cite{sorkin2005ten,bigatti2001tasi}. Indeed, the combination of quantum mechanics with gravity leads
	undoubtedly to a fuzzy picture of spacetime. A possible realization of this picture
	is offered by noncommutative and fuzzy geometry.
 
 The EE on fuzzy spaces was first studied in Refs.~\cite{Dou:2006ni,Dou:2009cw} for a free scalar field theory and was subsequently investigated by several authors \cite{Sabella-Garnier:2014fda,Karczmarek:2013jca,chen2018entanglement,okuno2016entanglement}, in particular for the fuzzy sphere. A central finding of these studies is that the free EE can, in general, be interpreted as being directly proportional to the number of degrees of freedom associated with the entangling boundary. This provides a natural realization of the area law in the fuzzy setting, with the usual UV divergence recovered in the commutative continuum limit. Moreover, it was shown that, despite the intrinsically nonlocal nature of the fuzzy-space description, the EE is dominated by degrees of freedom in the vicinity of the entangling boundary \cite{Dou:2009cw}, in close analogy with the behavior observed in continuum quantum field theories \cite{Das_2007,Das_2008}.
 
However, interacting noncommutative quantum field theories  exhibit a number of novel phenomena that are absent at the free level. Most notably, the celebrated UV/IR mixing phenomenon leads to IR singularities generated by UV loop momenta, with profound consequences for the renormalization structure of the theory and its low-energy effective description \cite{Minwalla_2000}. This phenomenon has also motivated deeper interpretations in terms of string theory, where UV/IR mixing can be understood as reflecting the extended nature of the underlying degrees of freedom. These features naturally raise the question of whether interactions can qualitatively modify the boundary-dominated structure of the entanglement entropy found at the free level, and, in particular, whether they can give rise to analogous UV/IR phenomena in the entanglement entropy.

It is worth noting that most studies of EE in the literature over the past three decades or so have focused on Gaussian theories, employing a variety of analytical and numerical techniques. These studies have addressed, among other issues, the structure of UV divergences, their universality across different critical models, and the extraction of finite and universal contributions to the EE, see for instance \cite{Nishioka:2018khk,Hertzberg:2010uv, Calabrese:2009qy,Casini:2009sr} and references therein. Nevertheless, a number of works in recent years have extended the study of EE to non-Gaussian interacting theories \cite{Hertzberg:2012mn,Chen:2020ild,Allouche_2022}. In particular, in \cite{Hertzberg:2012mn}, the EE associated with tracing over a half-space in $(d+1)$-dimensional spacetime was calculated perturbatively for scalar field theories with $\lambda\phi^4$ and $g\phi^3$ interactions. It was found that, in $d=3$, the leading-order result can be obtained from the free-theory expression by replacing the bare mass $m$ with the renormalized mass $m_r$ evaluated at the zero-momentum renormalization scale.

In this paper, we revisit the $2+1$-dimensional models—the fuzzy sphere and the fuzzy disc—studied in \cite{Dou:2006ni, Dou:2009cw}, go beyond the free Gaussian theory by introducing a $\lambda\phi^4$ interaction and compute its first-order contribution to  R\'{e}nyi entropies and EE. We employ the Green's function approach developed in \cite{Allouche:2018err, Allouche_2022} to construct a systematic perturbative expansion. The first-order correction is expressed in terms of the corresponding sectorial Green's functions and, after integration over Euclidean time, reduces to a functional sum over the spectrum of the fuzzy Laplacian. These sums are evaluated numerically for both geometries, allowing us to identify the leading scaling behavior and infer the corresponding analytic form of the leading contribution to the  R\'{e}nyi entropies, and consequently to the entanglement entropy.

In the fuzzy sphere case, the first-order correction exhibits a strong dependence on the mass and develops a pronounced IR divergence associated with the zero mode. We address this issue by projecting out the zero mode and introducing a corresponding more physically meaningful quantity. Our results reveal a clear departure from the free-theory behavior, and hence from the area law, in both models. The first-order interaction correction significantly modifies the structure of the free EE by introducing an extensive bulk factor that depends on the geometry of the entire fuzzy space. This stands in contrast to the leading free-theory EE, which was shown to be proportional to the number of boundary degrees of freedom and dominated by DF in the immediate vicinity of the entangling boundary.

We further investigate several relevant limits and find that the interaction correction develops an IR divergence in the Moyal plane limit. Several numerical investigations indicate that the extensive character of the interaction correction originates from the cumulative contributions of bulk DF: degrees of freedom in both the interior and exterior regions contribute significantly and with comparable weights. Finally, we address the origin of the IR divergence and its possible connection to the UV/IR mixing observed in noncommutative quantum field theories. This question is explored through a combination of numerical evidence and heuristic analytical arguments based on the spectral properties of the fuzzy Laplacian.

The paper is organized as follows. In Section 2, we present the general formalism used to compute perturbatively the corrections to the R\'{e}nyi entropies. Section 3 is devoted to a brief review of the main results for the free theories on fuzzy spaces, together with a specification of our conventions and notation. In Section 4, we derive the analytical form of the first-order correction to the R\'{e}nyi entropies for both models, introduce the zero-mode-projected Green's function for the fuzzy sphere, and present  the numerical calculations. Finally, in Section 5, we discuss our results in detail and examine their physical interpretation and possible implications.

 \section{The General Formalism}\label{section2}
 
 In this section, we briefly review the formalism used throughout this paper. For a more detailed derivation and discussion, we refer the reader to Ref.~\cite{Allouche_2022}.
 
For the models considered in this paper, it is sufficient to consider the following standard Lagrangian describing $N$ coupled harmonic oscillators with a quartic interaction: 
 \be\label{LagStand}
 L_E=\frac{1}{2}( \dot{Q}^T \dot{Q} +Q^T V Q+\lambda .Q.Q.Q.Q),~~ ~~~ Q=\left(
 \begin{array}{c}
 	q_1 \\
 	q_2 \\
 	\vdots\\
 	
 	q_N \\
 	
 \end{array}
 \right).
 \ee
 
 By $\lambda Q.Q.Q.Q$ we mean a generic quartic combination of the variables $q_i$, whose precise form will be specified later.
 
 Let us first consider the non-interacting case, $\lambda=0$. Let $\rho=|0\rangle\langle 0|$ denote the ground-state density operator of the full system, and consider the reduced density operator
 $\rho_A= \mathrm{Tr}_A \rho$,  obtained by tracing out a subset $\bar{A}$ of the field variables $Q$. 
 
 For the models considered in this work, $A$ and $\bar{A}$ will refer, respectively, to the following subsets:.

 $$
 A=\{ q_1,q_2,\cdots  q_p\}~~,~~A=\{ q_{p+1},q_{p+2},\cdots  q_N\},
 $$
 
 for some $p$.
 
 A natural decomposition of $Q$ then follows, namely
  \be
  Q=\left(
  \begin{array}{c}
  	Q_A \\ 	
  	Q_{\bar{A}} \\
  \end{array}
  \right), ~~~ Q_A=\left(
  \begin{array}{c}
  	q_1 \\ 	
  	q_2 \\
  	\vdots\\
  	q_p
  \end{array}
  \right),~~~Q_{\bar{A}}=\left(
  \begin{array}{c}
  	q_{p+1} \\ 	
  	q_{p+2} \\
  	\vdots\\
  	q_N
  \end{array}
  \right).
  \ee

 In  \cite{Allouche:2018err} it was shown that
  $ \mathrm{Tr}_{\bar{A}}\r_A^n$  with the correct normalization can be written as
 \be\label{partionf}
 \mathrm{Tr}_{\bar{A}}\r_A^n = \frac{\mathbb{Z}_n}{\mathbb{Z}_1^n}=\frac{\int \mathcal{D} \mathbb{Q}
 	e^{-\mathbb{S}_E}}{(\int \mathcal{D}Q e^{-S_E})^n}
 \ee
 
 and
  \be\label{id}
 \ln \mathrm{Tr} \rho_A^n= -\frac{1}{2} \ln \frac{\det (-\frac{d^2}{d\t^2}+\mathbb{V}(\t))}{\det
 	(-\frac{d^2}{d\t^2}+V)^n},~~~~-\infty <\t < \infty
 \ee
 where 
 \be\label{action}
 \mathbb{S}_E=\int_{-\infty}^{\infty} ( \frac{1}{2}[ \dot{\mathbb{Q}}^T \dot{\mathbb{Q}}
 +\mathbb{Q}^T\mathbb{ V(\t)} \mathbb{Q}] d\t, ~~~\mathbb{V} (\t)= \theta(-\t) \mathbb{V}+ \theta(\t)
 \mathbb{V}_p
 \ee
 where $\mathbb{Q}$ and $\mathbb{V}$ denote, respectively, the $n$-fold field variable and the corresponding $n$-fold potential, defined as follows:
 \be
 \mathbb{Q}= \left(
 \begin{array}{c}
 	Q^1_A \\
 	Q^1_{\bar{A}} \\
 	Q^2_A  \\
 	Q^2_{\bar{A}} \\
 	\cdot\\
 	\cdot\\
 	\cdot\\
 	
 	Q^n_A \\
 	Q^n_{\bar{A}}
 	
 \end{array}
 \right)
 ~~   \mathrm{and}~~~\mathbb{V}=
 \mathbb{I}_{n\times n}  \otimes  V
 \ee
 
 whereas $\mathbb{V}_p$ is  the transformed (permuted) $n$-fold potential given by
 
 \be
 \mathbb{V}_p= \mathbb{P}_\pi
 \mathbb{V}\mathbb{P}_\pi^T.
 \ee
   
  $\mathbb{P}_\pi$ is  $Nn\times Nn $ permutation matrix, its explicit form is given for the above particular choice of $A$  by
 \be\label{permutation}
 \mathbb{P}_{\pi}(A,\bar{A})=
 \left(
 \begin{array}{cccccc}
 	P_{A}& P_{\bar{A}} & 0_{N}&  \dots & 0_{N}&0_{N}\\
 	0_{N}& P_{A}&P_{\bar{A}}&  \dots  & 0_{N}& 0_{N}\\
 	0_{N}& 0_{N}& P_{A}& \dots & 0_{N}& 0_{N}\\
 	\vdots& \vdots& \vdots&\ddots & \vdots& \vdots\\
 	0_{N}& 0_{N}& 0_{N}& \dots & P_{A}& P_{\bar{A}}\\
 	P_{\bar{A}}& 0_{N}& 0_{N}& \dots & 0_{N}& P_{A}
 \end{array}
 \right)
 \ee
 
 $P_{A}$ and $P_{\bar{A}}$ are the projectors on $A$ and $\bar{A}$ respectively. More explicitly, we have
 
 $$
 P_A= \left(
 \begin{array}{cc}
 	I_p & 0\\
 	0& 0_{N-p}
 \end{array}
 \right), ~~~~~P_{\bar{A}}= \left(
 \begin{array}{cc}
 	0_p & 0\\
 	0&I_{N-p}
 \end{array}
 \right)
 ,~~~~
 $$
 
 It can then be shown that
 \be\label{Gh}
 \ln \mathrm{Tr}\rho_A^n=\frac{1}{2}\int_{0}^{\infty} (\mathrm{Tr} \mathbb{ G}_n(\t,\t,E)-n\mathrm{Tr} G(\t,\t,E)) dE
 \ee
 The trace $\mathrm{Tr}$ is understood to include integration over $\t$. $\mathbb{ G}_n(\t,\t,E) $ is the $n$-fold matrix Green's function associated  with  $-\frac{d^2}{d\t^2}+\mathbb{V}(\t) +E$  satisfying the following differential equation,
 
 \begin{equation}\label{GE}
 	(-\frac{d^2}{d\t^2}+\mathbb{V}(\t)+E)\mathbb{G}_n(\t,\t')=\delta(\t-\t') \mathbb{I},~~~~~G(\t,\t')=
 	\mathbb{G}_1(\t,\t').
 \end{equation}
 
 The explicit form of this $n$-fold Green's function for an arbitrary $\mathbb{V}(\t)$ is given in Ref.~\cite{Allouche:2018err} as follows:

  \be\label{GFS}
 \mathbb{G}_{\mp\mp}(\t,\t') = \frac{1}{2\mathbb{W}_\mp} e^{-\mathbb{W}_\mp|\t-\t'|}
 +\frac{1}{2\mathbb{W}_\mp}  e^{\pm\mathbb{W}_\mp\t}
 (\mathbb{W}_\mp-\mathbb{W}_\pm)(\mathbb{W}_-+\mathbb{W}_+)^{-1}   e^{\pm\mathbb{W}_\mp \t'}
 \ee
 
 and

 \be\label{GFS2}
 \mathbb{G}_{\pm\mp}(\t,\t')=  e^{\mp\mathbb{W}_\pm\t}(\mathbb{W}_-+\mathbb{W}_+)^{-1}
 e^{\pm\mathbb{W}_\mp \t'}
 \ee

 $$
 \frac{1}{\mathbb{W}_{\pm}}=\mathbb{W}_{\pm}^{-1}
 $$
 where $\mathbb{W}_{\pm}$ are given by
 
 $$
 \mathbb{W}_-=   \mathbb{I}_{n} \otimes \sqrt{(V+E)} ,~~~~\text{and}~~ \mathbb{W}_+=\mathbb{P}_\pi \mathbb{W}_- \mathbb{P}_\pi^T
 $$

 An alternative general expression for  R\'{e}nyi entropies of an arbitrary system of coupled harmonic oscillators can be readily derived from \eqref{Gh} and \eqref{GFS}. It takes the following form:

 \be\label{ent3}
 \ln \mathrm{Tr}\rho_A^n=  \frac{1}{8}\int_{0}^{\infty} \mathrm{Tr}  (
 \mathbb{V}_-^{-1}-\mathbb{V}_+^{-1})(\mathbb{W}_--\mathbb{W}_+)\mathbb{(W}_-+\mathbb{W}_+)^{-1} dE
 \ee
 
 Having derived the above expression for the R\'{e}nyi entropy using the Green's function approach, which is applicable to Gaussian theories, we are now in a position to go beyond the Gaussian approximation and incorporate corrections arising from quartic interactions.
 
 By construction, the Green's function approach provides a natural framework for developing a systematic perturbative expansion of the entanglement entropy in interacting theories.
 
 To this end, let us return to the original interacting model introduced above, namely the theory described by the Lagrangian \eqref{LagStand}. Following the same steps as in the free case, we can write:
 
 \be\label{partionfi}
 \mathrm{Tr}_{\bar{A}}\r_A^n(\l) = \frac{\mathbb{Z}_n(\l)}{\mathbb{Z}_1^n(\l)}=\frac{\int \mathcal{D} \mathbb{Q}
 	e^{-\mathbb{S}_E(\l)}}{(\int \mathcal{D}Q e^{-S_E(\l)})^n}
 \ee
 
 To the first order in $\l$ we can write
 
 \be\label{Perturb1}
 \ln \mathrm{Tr}\rho_A^n(\l)= \ln \mathbb{Z}_n(0)- n\ln \mathbb{Z}_1(0)-\l \int_{-\infty}^{\infty} d\t (< \mathbb{Q}.\mathbb{Q}.\mathbb{Q}.\mathbb{Q}>_0-n <Q.Q.Q.Q>_0)+\cdots
  \ee
   where 
  \be\label{Perturbnfold}
  < \mathbb{Q}.\mathbb{Q}.\mathbb{Q}.\mathbb{Q}>_0=\frac{\int \mathcal{D} \mathbb{Q}~\mathbb{Q}.\mathbb{Q}.\mathbb{Q}.\mathbb{Q} 
  	e^{-\mathbb{S}_E(0)}}{\int \mathcal{D}\mathbb{Q} e^{-\mathbb{S}_E(0)}}
  \ee

 and
 \be\label{Perturb1fold}
 <Q.Q.Q.Q>_0=\frac{\int \mathcal{D} Q~Q.Q.Q.Q 
 	e^{-\mathbb{S}_E(0)}}{\int \mathcal{D}Q e^{-S_E(0)}}
 \ee
 
 Now, using Wick's theorem together with the explicit form of the $n$-fold Green's function given in Eq.~\eqref{GFS}, we can obtain explicit analytic expressions for both $\langle \mathbb{Q}.\mathbb{Q}.\mathbb{Q}.\mathbb{Q}\rangle_0$ and $\langle Q.Q.Q.Q\rangle_0$.
 
 It is important to note that, up to this point, we have assumed that the interaction term $\l \mathbb{Q}.\mathbb{Q}.\mathbb{Q}.\mathbb{Q}$ is invariant under the action of the permutation matrix $\mathbb{P}_\pi$.
 
Before proceeding to apply Eq.~\eqref{Perturb1} to models based on fuzzy-space regularization, it is important to note that not every quartic interaction leads to a well-defined EE. For example, an interaction of the form \(\lambda (Q^T Q)^2\) gives rise to a pathological IR divergence that cannot be removed. Such IR-divergent interactions are generally understood to arise from the discretization of nonlocal interactions or higher-derivative terms. The absence of this type of IR divergence therefore restricts us to a specific class of quartic interactions; see Ref.~\cite{Allouche:2018err} for further discussion.

  \section{Entanglement Entropy on Fuzzy Spaces}
  
  Before tackling the computation of the correction to the EE due to quartic interactions in scalar field theories on $2+1$-dimensional fuzzy spaces, let us briefly review the main results for the corresponding free theories, thereby fixing our conventions and notation.
  
 We consider a scalar field of mass $m$ defined on $\mathbf{R}\times \mathbf{M}_N^2$, where $\mathbf{M}_N^2$ denotes a two-dimensional fuzzy space of matrix dimension $N$. We will consider two specific realizations: the fuzzy sphere $\mathbf{S}_N^2$ and the fuzzy disc $\mathbf{D}_N^2$.
 
 The action is given by

  \be\label{fuzzylag}
  S_0=\theta \int dt \frac{1}{2}Tr \big(\dot{\phi}^2
  +\phi\big[\Delta_N-m^2\big]\phi \big)
  \ee
  
 where $\theta =R^2/N$ is the noncommutativity parameter, $R$ is the radius of the fuzzy sphere or fuzzy disc, and $\Delta_N$ denotes the corresponding Laplacian.
  For the fuzzy sphere, the action of $\Delta_N$ on the matrix field is given by
  \be
  \Delta_N \phi = -\frac{{\cal L}_i^2}{R^2}(\phi)
  =-\frac{1}{R^2}[L_i,[L_i,\phi]] .
  \ee
  
  The $L_i$ generate the $SU(2)$ irreducible representation of spin $l=\frac{N-1}{2}$.
  
  For the fuzzy disc, the action of $\Delta_N$ is given by 
  \be
  \Delta_N \phi = -\frac{4}{\theta^2}[a^+,[a,\phi]], ~~~~[a,a^+]=\theta .
  \ee
  
  The fuzzy disc is obtained by truncating the algebra of the Moyal plane to finite-dimensional $N\times N$ matrices $\phi$, namely,
  \be
  \phi =\sum_{n,m=0}^{N-1}\phi_{mn}|m\rangle\langle n|, ~~~\phi^*_{nm}=\phi_{mn}.
  \ee
  
  The Laplacian $\Delta_N$ then has a well-defined action on $\phi$ Ref.~\cite{Lizzi_2003}. The radius of the fuzzy disc is given by
  \be
  R^2=N\theta .
  \ee
  
To bring the action \eqref{fuzzylag} into the canonical form \eqref{LagStand}, one works for the fuzzy sphere in the basis in which $L_3$ is diagonal and introduces suitable field variables, as described in \cite{Dou:2006ni}. In this basis, the free scalar theory on the fuzzy sphere naturally decomposes into $2(2l)+1=2N-1$ independent sectors, ${\mathcal{H}_m}$, with
 $$
\{\mathcal{H}_m\} , m=-(N-1),\cdot\cdot\cdot\cdot, (N-1).
$$
Each sector $\mathcal{H}_m$ contains $N-|m|$ degrees of freedom, corresponding to $N-|m|$ coupled harmonic oscillators, and is described by a Lagrangian $L^{(m)}$. More explicitly, we have
\be
L=\sum_{-(N-1)}^{N-1}L^{(m)}=
\sum_{m=-(N-1)}^{N-1}\frac{1}{2}[\dot{Q}_m^T \dot{Q}_m-Q_m^TV^{(m)} Q_m]
\ee

where
\begin{eqnarray}\label{fspotential}
	V_{ab}^{(m)}=\frac{2}{R^2}\bigg[\big(c_2+\frac{{\mu}^2}{2}-A_aA_{a+|m|}\big){\delta}_{a,b}
	-\frac{1}{2}B_{a-1}B_{a-1+|m|}{\delta}_{a-1,b}-\frac{1}{2}B_{a}B_{a+|m|}{\delta}
	_{a+1,b}\bigg]
\end{eqnarray}

 $ B_a = \sqrt{a(N-a)}$ and $ A_a =-a+\frac{N+1}{2}$,  $a,b= 1\dots  N-|m|$, and $\mu=mR$ is the dimensionless mass parameter.

The new canonical variables $Q_a^{(m)}$ arise naturally after decomposing $\phi$ as $\phi=\phi_1+i\phi_2$ and defining 

\be\label{Phi}
\Phi=  \phi_1+\phi_2.
\ee

\begin{eqnarray}\label{cv1}
	Q^{(m)}=\left(
	\begin{array}{c}
		Q_1^{(m)} \\
		Q_2^{(m)} \\
		\vdots\\
		
		Q_{N-m}^{(m)} \\	
	\end{array}
	\right)=\sqrt{\theta}\left(
	\begin{array}{c}
	\Phi^{1,1+m} \\
		\Phi^{2,2+m} \\
		\vdots\\
		
		\Phi^{N-m,N} \\	
	\end{array}
	\right),~~~ m\ge 0.
\end{eqnarray}
\begin{eqnarray}\label{cv2}
	Q^{(m)}=\left(
	\begin{array}{c}
		Q_1^{(m)} \\
		Q_2^{(m)} \\
		\vdots\\
		
		Q_{N-|m|}^{(m)} \\	
	\end{array}
	\right)=\sqrt{\theta}\left(
	\begin{array}{c}
		\Phi^{1+|m|,1} \\
		\Phi^{2+|m|,2} \\
		\vdots\\
		
		\Phi^{N,N-|m|} \\	
	\end{array}
	\right),~~~ m\le 0.
\end{eqnarray}

To introduce the EE in these fuzzy-space models, one must first divide the field degrees of freedom into two or more disjoint regions. In \cite{Dou:2006ni}, it was conjectured, based on heuristic arguments, that the field variables represented by the elements of the upper-left triangular part of the field matrix $\Phi_R$ and those in the lower-right triangular part $\Phi_L$ correspond, respectively, to the upper and lower fuzzy hemispheres. This conjecture was subsequently established rigorously in \cite{Karczmarek:2013jca} using mathematical arguments.
More explicitly for $N=5$ we have

\begin{eqnarray}
	\Phi_{L}=\left (\begin{array}{ccccc}
		{\Phi}_{11} & {\Phi}_{12} & {\Phi}_{13} & {\Phi}_{14} & 0 \\
		{\Phi}_{21} & {\Phi}_{22} & {\Phi}_{23} & 0 & 0 \\
		{\Phi}_{31} & {\Phi}_{32} & 0 & 0 & 0 \\
		{\Phi}_{41} & 0 & 0 & 0 & 0 \\
		0 & 0 & 0 & 0 & 0
	\end{array}\right)~,~
	\Phi_{R}=\left (\begin{array}{ccccc}
		0& 0 & 0 & 0 & \Phi_{15}\\
		0 &0 & 0 & \Phi_{24}  &  \Phi_{25}\\
		0 & 0 & \phi_{33}  & \phi_{34} & \Phi_{35}  \\
		0 &  \Phi_{42} & \Phi_{43} & \Phi_{44}  & \Phi_{45} \\
		\Phi_{51}& \Phi_{52} & \Phi_{53} & \Phi_{54} &  \Phi_{55}
	\end{array}\right)
\end{eqnarray}
\begin{figure}
	\centering
	\includegraphics[width=0.9\textwidth]{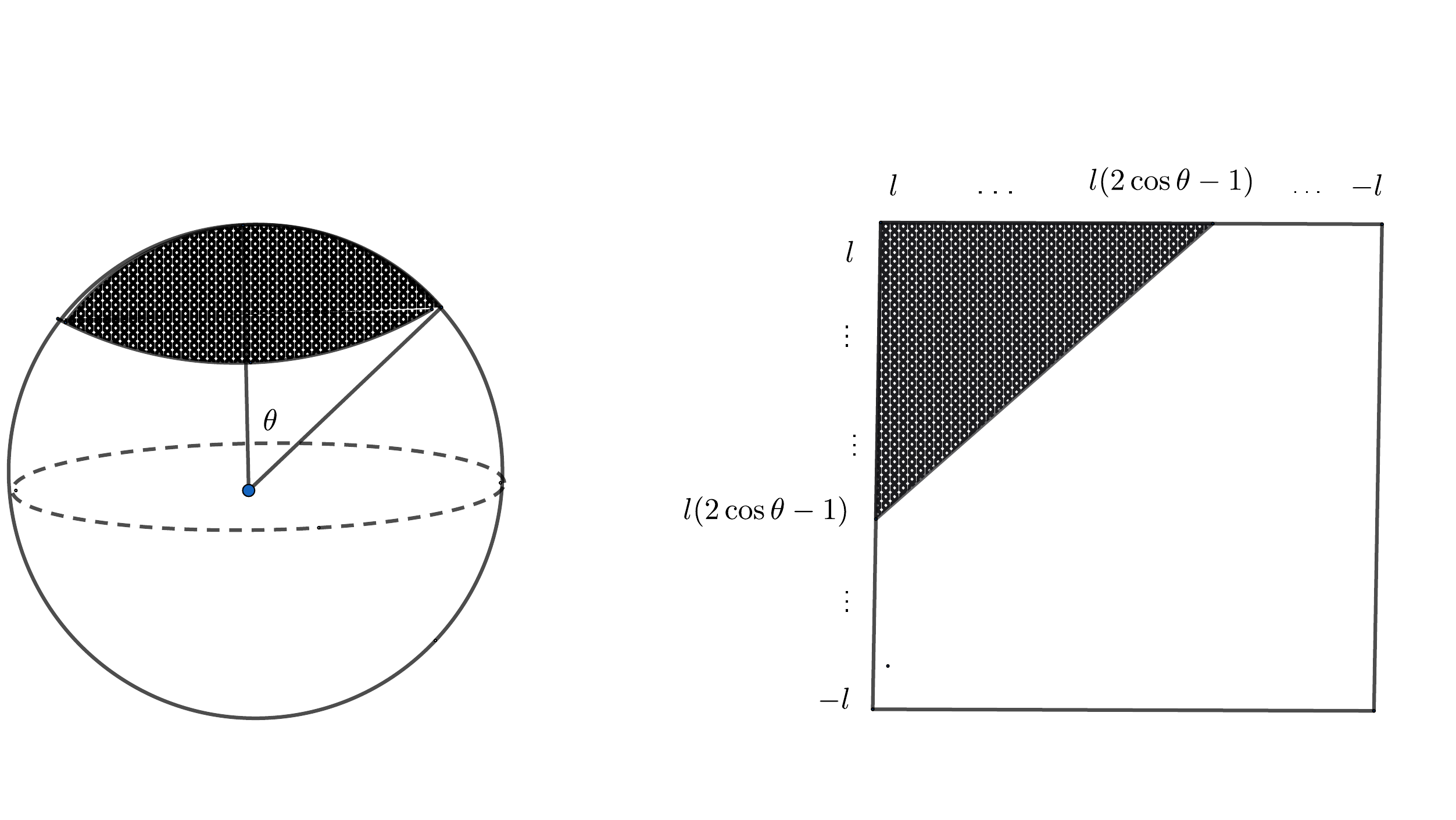}
	\caption{Degrees of freedom on the sphere and their matrix counterparts.}
	\label{figFvC}
\end{figure}

More generally, it was shown in Ref.~\cite{Karczmarek:2013jca} that the DF corresponding to a polar cap $C$ of angular radius $\theta$, as illustrated in Figure~\ref{figFvC}, can be identified, in the large-$l$ (or equivalently large-$N$) limit, with the set of matrix elements $\langle m_1|\phi|m_2\rangle$ satisfying $m_1 + m_2 > 2l \cos\theta$,  where $l=\frac{N-1}{2}$ \footnote{Here, the eigenvalues of $L_z$ are used to label the matrix elements.}. Although computing the EE for a hemisphere requires only the condition $m_1+m_2>0$, this more general result will be useful later when we identify the degrees of freedom contributing to the EE and, in particular, to its interaction correction.

In terms of the canonical field variables $Q_m$, tracing out either the upper or the lower fuzzy hemisphere corresponds to tracing out half of the degrees of freedom in each individual sector $\mathcal{H}_m$.

Since the different sectors do not mix in the free theory, the computation of the EE reduces to computing the entropy contribution from each sector separately. Each sector is described by a Lagrangian of the standard form \eqref{LagStand}. Consequently, the ground-state density operator of the full system factorizes as
\begin{eqnarray}
	\rho=\bigotimes_{m=-(N-1)}^{N-1} \rho^{(m)}.
\end{eqnarray}

and the corresponding reduced density operator, obtained by tracing out the lower hemisphere, $\rho_{R}=\mathrm{Tr}_{L}\rho$, is given by
\begin{eqnarray}
	\rho_{R}=\bigotimes_{m=-(N-1)}^{N-1} \rho_R^{(m)}.
\end{eqnarray}

where $\rho_R^{(m)}$ is obtained by tracing out, within each sector, the degrees of freedom corresponding to the lower hemi-fuzzy sphere.

In \cite{Dou:2006ni}, it was shown numerically, using the real-time approach, that the resulting EE is proportional to the number of degrees of freedom associated with the entangling boundary. This result can equivalently be interpreted as an area law, with the EE proportional to the circumference of the separating region, namely, the fuzzy circle. More explicitly, we have
\be
S_{ent}= 0.39 N+\cdots
\ee

where the ellipsis denotes subleading, mass-dependent terms, which most likely have a logarithmic dependence on the mass.

For the fuzzy disc, a similar decomposition of the matrix field \(\phi\) in the oscillator basis brings the action into the following form:

 \be
 S=\int L dt,  ~~~~ 
 L=\sum_{-(N-1)}^{N-1}L^{(m)}=
 \sum_{m=-(N-1)}^{N-1}\frac{1}{2}[\dot{Q}^{(m)T} \dot{Q}^{(m)}-{Q}^{(m)T}V^{(m)} {Q}^{(m)T}]
 \ee
 
 where $V^{(m)}$ is given by

\begin{eqnarray}\label{fdpotential}
	V_{ab}^{(m)}=\frac{2}{\theta}\bigg[\big(2a +|m|-1+\frac{{\mu}^2}{4}\big){\delta}_{a,b}
	&-&\sqrt{(a-1)(a-1+|m|)}{\delta}_{a-1,b}\nonumber \\  -\sqrt{a(a+|m|)}{\delta}
	_{a+1,b}\bigg], ~~\mu=m\sqrt{\theta}.
\end{eqnarray}

For the fuzzy disc, two different prescriptions for tracing out the DF were studied in \cite{Dou:2006ni, Dou:2009cw}. Here, however, we shall reconsider only one of them, namely, the case in which we trace out the DF residing in a smaller sub-disc. This choice is particularly natural in the oscillator basis, where the radial bipartition of the degrees of freedom is directly related to the structure of the basis.

The degrees of freedom of the sub-disc $D_q$ to be traced out are represented by the following $q\times q$ submatrix, with $q<N$:

\begin{eqnarray}\label{subdiscmatrix} 
	\Phi_{q}=\left (\begin{array}{cccc} \phi_{11}& \phi_{12} & \cdots & \phi_{1q}\\ \phi_{21} &\phi_{22} & \cdots & \phi_{2q} \\ \cdots & \cdots &\cdots & \cdots \\ \phi_{q1} & \phi_{q2} & \cdots& \phi_{qq} \\ \end{array}\right) 
\end{eqnarray}

As in the fuzzy-sphere case, the computation of the EE can be carried out separately in each angular-momentum sector, with the total EE obtained by summing the individual sector contributions. For the sub-disc bipartition, however, not all angular sectors contribute to the EE: only the DF belonging to sectors with $|m|<q$ are entangled with the DF of the sub-disc. The sectors with $|m|\ge q$ are unentangled with $D_q$ and therefore give no contribution to the free-theory EE.

Again, it was found that the EE is proportional to the number of degrees of freedom associated with the entangling boundary. More explicitly, one obtains
\be
S_{ent}(q)=0.46q, ~~~\text{for}~ N, q\gg 1.
\ee

\section{Interaction Correction to the Entanglement Entropy}

Having devoted the two previous sections to laying out the general formalism of the Green's function approach and reviewing the models considered in this work, we are now in a position to compute the first-order correction to the R\'{e}nyi entropy arising from the quartic interaction in scalar field theories on $2+1$-dimensional fuzzy spaces. We consider two cases in turn: first, a fuzzy sphere and then a fuzzy disc.
Let us add a quartic interaction term to the action \eqref{fuzzylag},
\be\label{intaction}
S= \theta \int dt \bigg[\frac{1}{2} \mathrm{Tr} \big(\dot{\phi}^2
+\phi\big[\Delta_N-m^2\big]\phi \big) -\frac{\l}{4!} \mathrm{Tr} \phi^4\bigg].
\ee

As discussed above, the free (quadratic) part of the Lagrangian naturally decomposes into $2N-1$ uncoupled sectors, ${\mathcal{H}_m}$, with $m=-(N-1),\ldots,(N-1)$. The quartic interaction, however, couples these otherwise independent sectors,\footnote{This mixing is analogous to the coupling between different Fourier (momentum) modes induced by quartic interactions in standard continuum field theory.} thereby destroying the sector-by-sector decoupling . Nevertheless, the general formalism developed in Section~\ref{section2} remains applicable to these models in a rather straightforward way.

To this end, we arrange the canonical variables $Q^{(m)}$ into a single column vector with $2N-1$ components, namely,

\begin{eqnarray}\label{cv3}
	Q=\left(
	\begin{array}{c}
		Q^{-(N-1)} \\
		Q^{-(N-2)} \\
		\vdots\\
		Q^{(0)}\\
		\vdots\\
		Q^{(N-2)}\\
		Q^{(N-1)} \\	
	\end{array}
	\right)
\end{eqnarray}
Then the action \eqref{intaction} takes the following form,\footnote{We have switched to Euclidean time.}
\be\label{intaction2}
S_E= \int d\t \bigg[\frac{1}{2} \big(\dot{Q}^T\dot{Q}
+Q^TVQ \big) +\l Q.Q.Q.Q\bigg],
\ee
where $V$ is given by
\be\label{fullpotential}
V=\bigoplus_{m=-(N-1)}^{N-1} V^{(m)}.
\ee

The matrices $V^{(m)}$ represent the potentials associated with the individual sectors $m$; their matrix elements are given by \eqref{fspotential} and \eqref{fdpotential} for the fuzzy sphere and the fuzzy disc, respectively.

We keep the interaction term in a formal form, which is sufficient for capturing the general structure of the models considered in this paper.

Using \eqref{Perturb1}, \eqref{Perturbnfold}, and \eqref{Perturb1fold}, the first-order correction to $\ln \mathrm{Tr} \rho_A^n(\l)$ can be written formally as

\be
\delta \ln \mathrm{Tr} \r_A^n= -\l \int_{-\infty}^{\infty} d\t (< \mathbb{Q}.\mathbb{Q}.\mathbb{Q}.\mathbb{Q}>_0-n <Q.Q.Q.Q>_0) 
\ee

where $<\mathbb{Q}.\mathbb{Q}.\mathbb{Q}.\mathbb{Q}>_0$ and $<Q.Q.Q.Q>_0$ are evaluated using Wick's theorem and the corresponding Green's function matrices. More explicitly, we have

\begin{equation}
	\wick{\c {\mathbb{Q}}_i \c {\mathbb{Q}}_j} =  \mathbb{G}_{ij}(\t ,\t),~~~\wick{\c {Q}_i \c {Q}_j} =  G_{ij}(\t ,\t)
\end{equation}

 Where $\mathbb{G}_{ij}(\t.\t)$ and $G_{ij}(\t ,\t)$ are given by \eqref{GFS}. 
 
 In view of \eqref{fullpotential} we can write

 \be\label{greenff}
 \mathbb{ G}_n(\t,\t')= \bigoplus_{m=-(N-1)}^{N-1} \mathbb{ G}^{(m)}_n(\t,\t')
 \ee
While the arrangement of the DF in the column vector \eqref{cv3} shows that, in principle, one can construct an explicit expression for the first-order corrections and evaluate them numerically, this approach is not particularly efficient in practice. A more practical and efficient decomposition of the DF, tailored to the specific structure of the models considered here, will be introduced below.

\subsection{Interaction Correction on The Fuzzy Sphere}

Consider the interaction term appearing in \eqref{intaction},
\be\label{VintS}
V_{int}= \frac{\lambda}{4!} \mathrm{Tr} (\phi^4).
\ee

In order to compute the first-order correction using the Green's function technique developed in the previous section, together with the explicit form of the Green's function given by \eqref{GFS}, we first need to express the interaction explicitly in terms of the canonical variables $Q^{(m)}$.

We begin by expanding the interaction in terms of the field $\Phi$ defined in \eqref{Phi}, which gives
\be\label{traceint}
\mathrm{Tr}\phi^4= -\frac{1}{2}\mathrm{Tr}\big[\Phi^4- (\Phi\Phi^\dagger)^2 -2\Phi^2\Phi^{\dagger 2}\big].
\ee

To proceed and express the above trace in terms of the variables $Q^{(m)}$, it is useful to decompose the matrix field $\Phi$ into its upper- and lower-triangular parts, $ \Phi=\Phi^++\Phi^-$. For example, for $N=5$,

\begin{eqnarray}
	\Phi^{+}=\left (\begin{array}{ccccc}
		{\Phi}_{11} & {\Phi}_{12} & {\Phi}_{13} & {\Phi}_{14} & {\Phi}_{15} \\
		0& {\Phi}_{22} & {\Phi}_{23} & {\Phi}_{24} & {\Phi}_{25} \\
		0& 0&{\Phi}_{33} & {\Phi}_{34} & {\Phi}_{35} \\
		0& 0 &  0&{\Phi}_{44} & {\Phi}_{45} \\
		0& 0& 0& 0& {\Phi}_{55}
	\end{array}\right)~,~
	\Phi^{-}=\left (\begin{array}{ccccc}
		0& 0 & 0 & 0 & 0\\
		\Phi_{21}  & 0& 0& 0& 0\\
		\phi_{31}  & \phi_{32} & 0& 0& 0 \\
		\Phi_{41} & \Phi_{42} & \Phi_{43}  & 0& 0 \\
		\Phi_{51}& \Phi_{52} & \Phi_{53} & \Phi_{54} &  0
	\end{array}\right)
\end{eqnarray}

The main advantage of introducing $\Phi^+$ and $\Phi^-$ is that $\Phi^+$ contains exclusively the variables ${Q^{(m)}}$ with positive $m$, whereas $\Phi^-$ contains only the variables ${Q^{(m)}}$ with strictly negative $m$. Moreover, the matrix elements of $\Phi^+$ and $\Phi^-$ can be expressed directly in terms of the variables ${Q^{(m)}}$. Explicitly, we have

\be
\Phi^+_{ij}=Q^{(j-i)}_i,\ \Phi_{ij}^-=Q^{(i-j)}_j.
\ee

Here, for the time being, we have omitted the factor $\sqrt{\theta}$ appearing in the definition of $Q$. All factors of $\sqrt{\theta}$ can be straightforwardly restored at the end.

Furthermore, in view of the form of the Green's function, \eqref{greenff}, all Wick contractions between variables ${Q^{(m)}}$ belonging to different sectors $m$ are identically zero. Therefore, when expanding \eqref{traceint} in terms of $\Phi^+$ and $\Phi^-$, only terms containing even powers of $\Phi^+$ and $\Phi^-$ can give non-vanishing contributions.

Let us now illustrate how to calculate $<\mathrm{Tr} \phi^4>_0$ and express it in terms of the components of the Green's function matrix. Since the calculation is rather lengthy, we shall restrict ourselves to showing explicitly how one of the terms in the expansion \eqref{traceint} is evaluated. The remaining terms can be treated in exactly the same way, and we shall simply quote the final result.

Consider, as an example, the term $\mathrm{Tr}(\Phi^4)$ for $\t<0$.

\begin{eqnarray}
	\mathrm{Tr}(\Phi^4)=\mathrm{Tr}(\Phi^{+4}+\Phi^{-4}+4\Phi^{-2}\Phi^{+2}+4\Phi^{+3}\Phi^{-}+4\Phi^{-3}\Phi^{+}+2(\Phi^{+}\Phi^{-})^2)
\end{eqnarray}
In view of the above discussion, only terms containing even powers of $\Phi^+$ survive.
\begin{eqnarray}\label{expanterm1}
	\mathrm{Tr}(\Phi^4)=\mathrm{Tr}(\Phi^{+4}+4\Phi^{-2}\Phi^{+2}+2(\Phi^{-}\Phi^{+})^2)
\end{eqnarray}
Let us focus, as an example, on the term $\mathrm{Tr}(\Phi^{-2}\Phi^{+2})$. It is straightforward to see that, for the 1-fold field, we have
\begin{eqnarray}\label{term1-fold}
	\mathrm{Tr}(\Phi^{-2}\Phi^{+2})&=&\sum_{i,j,k,l} \Phi_{ij}^-\Phi_{jk}^-\Phi_{kl}^+\Phi_{li}^+\nonumber\\
	&=& \sum_{i>j>k,k\leq l\leq i} Q^{(j-i)}_jQ^{(k-j)}_kQ^{(l-k)}_kQ^{(i-l)}_l
\end{eqnarray}

 For $n$-fold version of 	$\mathrm{Tr}(\Phi^{-2}\Phi^{+2})$,  we have,

\begin{eqnarray}\label{termn-fold}
	\mathrm{Tr}(\mathbf{\Phi}^{-2}\mathbf{\Phi}^{+2})=\sum_{p=0}^{n-1}\sum_{\{i,j,k,l\}} \mathbb{Q}^{(j-i)}_{[j]p}\mathbb{Q}^{(k-j)}_{[k]p}\mathbb{Q}^{(l-k)}_{[k]p}\mathbb{Q}^{(i-l)}_{[l]p}\label{eqt:8}.
\end{eqnarray}
The indices ${i,j,k,l}$ run from $1$ to $N$ and are subject to the conditions ${i>j>k,k\leq l\leq i}$. The abbreviated notation $\mathbb{Q}^{(m)}_{[i]p}$ is defined as follows:
$$
\mathbb{Q}^{(m)}_{[i]p}\equiv \mathbb{Q}^{(m)}_{i+p(N-|m|)}
$$

Applying Wick's theorem to Eqs.~\eqref{term1-fold} and \eqref{termn-fold}, we obtain,
\begin{eqnarray}
	<\mathrm{Tr}({\Phi}^{-2}{\Phi}^{+2})>_0=	\sum_{i=2}^{N-1}\sum_{j=i+1}^{min(N,2i-1)}\mathbb{G}^{(i-j)}_{i,2i-j}\mathbb{G}^{(j-i)}_{2i-j,i}
\end{eqnarray}
\begin{eqnarray}
<\mathrm{Tr}(\mathbf{\Phi}^{-2}\mathbf{\Phi}^{+2})>_0=	\sum_{p=0}^{n-1}\sum_{i=2}^{N-1}\sum_{j=i+1}^{min(N,2i-1)}\mathbb{G}^{(i-j)}_{[i,2i-j]p}\mathbb{G}^{(j-i)}_{[2i-j,i]p}\label{eqt:9}
\end{eqnarray}
where 
$$
\mathbb{G}^{(m)}_{[ij]p}\equiv \mathbb{G}^{(m)}_{i+p(N-m),j+p(N-m)}
$$

On the other hand, it is not difficult to show that, within the present Green's function formulation, the $nN\times nN$ Green's function is an $N\times N$ block-circulant matrix.\footnote{This follows from the fact that the permutation matrix $\mathcal{P}_\pi$ introduced in \eqref{permutation} is itself an $N\times N$ block-circulant matrix.} Therefore, it follows, for instance, that

$$
\sum_{p=0}^{n-1}\mathbb{G}^{(m)}_{[ij]p}=n\mathbb{G}^{(m)}_{ij}
$$

Moreover, since $\mathbb{G}(\tau,\tau)$ is a Hermitian matrix and $\mathbb{G}^{(m)}=\mathbb{G}^{(-m)}$, Eq.~\eqref{eqt:9} takes the following simpler form:
\begin{eqnarray}\label{expectation1}
<\mathrm{Tr}(\mathbf{\Phi}^{-2}\mathbf{\Phi}^{+2})>_0=	n\sum_{i=2}^{N-1}\sum_{j=i+1}^{min(N,2i-1)}\big(\mathbb{G}^{(j-i)}_{i,2i-j}\big)^2
\end{eqnarray}
Similarly, one obtains for the other terms in the expansion \eqref{expanterm1}
\begin{eqnarray}
	<\mathrm{Tr}(\mathbf{\Phi}^{-}\mathbf{\Phi}^{+})^2>_0&=& n\sum_{i=2}^{N}\sum_{j=1}^{i-1}\big(\mathbb{G}^{(i-j)}_{jj}\big)^2\\
	<\mathrm{Tr}(\mathbf{\Phi}^{+4})>_0&=& 3n\sum_{i=1}^{N}\big(\mathbb{G}^{(0)}_{ii}\big)^2
\end{eqnarray} 
Therefore we obtain for $\mathrm{Tr}(\mathbf{\Phi}^4)$,
\begin{eqnarray}
	<\mathrm{Tr}(\mathbf{\Phi}^4)>_0&=& 3n\sum_{i=1}^{N}\big(\mathbb{G}^{(0)}_{ii}\big)^2+4n\sum_{i=2}^{N-1}\sum_{j=i+1}^{min(N,2i-1)}\big(\mathbb{G}^{(j-i)}_{i,2i-j}\big)^2+\nonumber\\
	&&+2n\sum_{i=2}^{N}\sum_{j=1}^{i-1}\big(\mathbb{G}^{(i-j)}_{jj}\big)^2\label{eqt:14}
\end{eqnarray}
The remaining terms in the interaction \eqref{traceint} can be evaluated in the same manner, yielding, for $\t<0$,

\begin{eqnarray}
<\mathbb{V}_{int}>_0|_{\t<0}
	&=&\frac{n}{2}\frac{\l}{4!}\Big[ 6\sum_{i=1}^{N}\big(\mathbb{G}^{(0)}_{ii}\big)^2+4 \sum_{i=1}^{N-1}\sum_{j=i+1}^{N}\mathbb{G}_{jj}^{(0)}\mathbb{G}^{(j-i)}_{ii}+
8\sum_{i=1}^{N-1}\sum_{j=i}^{N-1}\sum_{k=j+1}^{N}\mathbb{G}_{ii}^{(j-i)}\mathbb{G}_{jj}^{(k-j)}\nonumber\\  &+&4\sum_{i=2}^{N}\sum_{j=1}^{i-1}\sum_{k=j+1}^N\mathbb{G}_{jj}^{(i-j)}\mathbb{G}_{jj}^{(k-j)}+
	4\sum_{i=2}^{N}\sum_{j=1}^{i}\sum_{k=1}^{i-1}\mathbb{G}_{jj}^{(i-j)}\mathbb{G}_{kk}^{(i-k)}\nonumber\\&+& 8\sum_{i=2}^{N}\sum_{j=i}^{N}\sum_{k=1}^{i-1}\mathbb{G}_{ik}^{(j-i)}\mathbb{G}_{k,k+j-i}^{(i-k)}\Big]
\end{eqnarray}

Which can be recapitulated  as

\begin{eqnarray}
	<\mathbb{V}_{int}>_0|_{\t<0}
	&=&\frac{n}{2}\frac{\l N R}{4!}\Big[6\sum_{i=1}^{N}\big(\mathbb{G}^{(0)}_{ii}\big)^2+4 \sum_{l=1}^{N-1}\sum_{i=1}^{N-l}\mathbb{G}_{l+i,l+i}^{(0)}\mathbb{G}^{(l)}_{ii}\nonumber\\
	&+&8\sum_{m=1}^{N-1}\sum_{l=0}^{N-m-1}\sum_{i=1}^{N-m-l}\mathbb{G}_{ii}^{(l)}\mathbb{G}_{l+i,l+i}^{(m)}+\nonumber
\end{eqnarray}
	\begin{eqnarray}\label{cotractiontnegative}
		&+&4\big(\ \sum_{m=1}^{N-1}\sum_{l=1}^{m}\sum_{i=1}^{N-m}\mathbb{G}_{ii}^{(m)}\mathbb{G}_{ii}^{(l)}
	+\sum_{m=1}^{N-2}\sum_{l=m+1}^{N-1}\sum_{i=1}^{N-l}\mathbb{G}_{ii}^{(m)}\mathbb{G}_{ii}^{(l)} \big)\nonumber\\	
	&+&4\big(\sum_{m=1}^{N-1}\sum_{l=0}^{m}\sum_{i=m+1}^{N}\mathbb{G}_{i-l,i-l}^{(l)}\mathbb{G}_{i-m,i-m}^{(m)}
	+\sum_{m=1}^{N-2}\sum_{l=m+1}^{N-1}\sum_{i=l+1}^{N}\mathbb{G}_{i-l,i-l}^{(l)}\mathbb{G}_{i-m,i-m}^{(m)}\big)
	\nonumber\\
	&+&8\sum_{m=0}^{N-2}\sum_{l=1}^{N-m-1}\sum_{i=l+1}^{N-m}\mathbb{G}_{i,i-l}^{(m)}\mathbb{G}_{i-l,i-l+m}^{(l)}\Big].
\end{eqnarray}

where we have restored the correct canonical normalization, and $\mathbb{G}^{(m)}$ are defined by \eqref{GFS} with the rescaled potentials $V^{(m)}\rightarrow R^2 V^{(m)}$.

Having calculated the contributions for $\t<0$, we now need to evaluate $<\mathbb{V}_{int}>_0$ for $\t>0$, or equivalently, $<\mathbb{P}_\pi\mathbb{V}_{int}\mathbb{P}_\pi^T>_0$.

Let us first note that the interaction potential is not invariant under the action of the permutation $\mathbb{P}_\pi$. Nevertheless, it is not necessary to evaluate the $\t>0$ contribution separately. Indeed, it can be shown that, after integration over $\t$, the contribution from $\t>0$ is identical to that from $\t<0$. More explicitly, we have,

\be
\int_0^\infty <\mathbb{P}_\pi\mathbb{V}_{int}\mathbb{P}_\pi^T>_0 d\t =\int_{-\infty}^0 <\mathbb{V}_{int}>_0 d\t
\ee

We can now write the first-order correction to $\ln  \mathrm{Tr}\rho_R^n$ as

\begin{eqnarray}\label{correctionf1}
\delta \ln \mathrm{Tr}\r_R^n&=& -\int_{-\infty}^\infty \big(<\mathbb{V}_{int}>_0 -n<V_{int}>_0\big )d\t\nonumber \\
&=& -2\int_{-\infty}^0 \big(<\mathbb{V}_{int}>_0 -n<V_{int}>_0\big )d\t
\end{eqnarray}

where $<{V}_{int}>_0$ is obtained from \eqref{cotractiontnegative} simply by replacing $\mathbb{G}^{(m)}$ with the corresponding 1-fold Green's functions $G^{(m)}$.

To perform the integration over Euclidean time, we use the spectral decompositions of the matrices $\mathbb{W}^{(m)}$. The resulting analytical expression is rather lengthy and cumbersome and is therefore given in Appendix~\ref{Appendix I}. Moreover, evaluating the resulting multiple sums analytically in the large-$N$ limit appears to be intractable. We therefore evaluate the sums numerically for different values of $N$, the mass, and $n$.
To this end, we define the following rescaled dimensionless function:
\be\label{fnmuN}
f_n(\mu,N)=\frac{4!}{\l NR}\frac{\delta \ln Tr \r_R^n}{(n-1)}.
\ee

We then evaluate $f_n(\mu,N)$ numerically using the explicit expression given in  Appendix~\ref{Appendix I}.

%

\begin{figure}[h]
	\centering
	\includegraphics[trim=0cm 8cm 0cm 8cm, clip, width=0.9\linewidth]{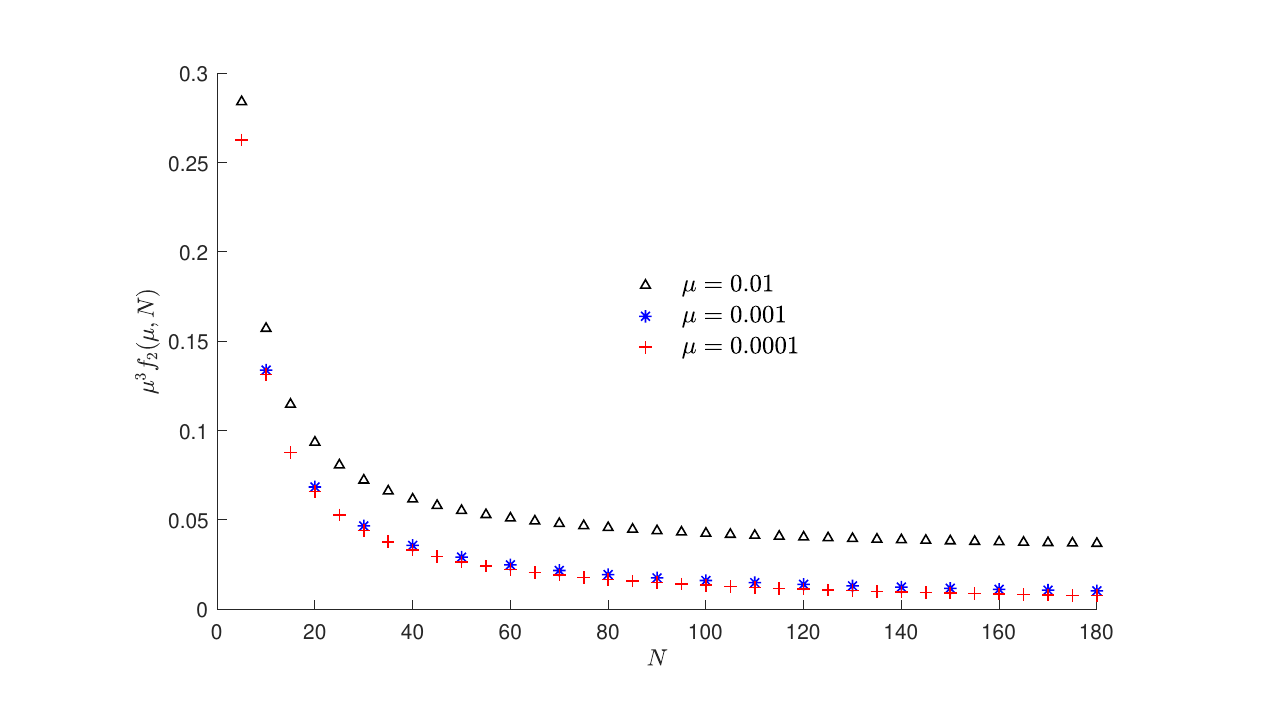}
	\caption{The rescaled dimensionless function $\mu^3f_n(\mu,N)$ as a function of $N$, for different values of the mass parameters and $n=2$.}
	\label{FigZM}
\end{figure}

Figure~\ref{FigZM} shows $\mu^3 f_n(\mu,N)$ for different values of the mass parameter, $\mu=10^{-2},10^{-3},10^{-4}$, with $n=2$, as a function of $N$.

The first notable feature is the strong dependence of the correction on the mass. In contrast to the free (Gaussian) EE, which is predominantly UV dominated, the first-order interaction correction obtained within naive perturbation theory exhibits a pronounced dependence on the mass, indicating a strong sensitivity to the IR sector. Our numerical results strongly suggest the following approximate large-$N$ scaling for the first-order correction:

\be
\delta \ln \mathrm{Tr} \r_R^n \sim \lambda (n-1) \frac{NR}{\mu^3}  .
\ee

This mass scaling is not surprising. As can be seen from the expression given in the appendix, it can be traced to the lowest eigenvalue of $V^{(0)}$, which approaches zero in the massless limit and corresponds to the zero mode of the Laplacian on the fuzzy sphere.

In contrast to the continuum flat-space case studied in Ref.~\cite{Hertzberg:2012mn}, where the zero mode has measure zero and is effectively regulated by the infinite spatial volume, the fuzzy sphere describes a finite-volume quantum field theory in which the zero mode acts as a genuine collective coordinate. The resulting IR divergence therefore signals a breakdown of naive perturbation theory: as the mass becomes small, the quartic correction to the EE becomes increasingly sensitive to the zero mode and eventually becomes non-perturbative. Consequently, an expansion in small $\lambda$ ceases to be reliable in this regime, since perturbation theory is effectively being performed around an inappropriate Gaussian vacuum.

Therefore, before any reliable perturbative result can be extracted within our approach, the zero mode must be treated consistently.

In general, there are two ways to deal with the zero-mode problem. The first, and technically more challenging, is to treat the zero mode non-perturbatively as a collective coordinate, while carrying out perturbation theory only around the nonzero modes. The second is to project out the zero mode by imposing an appropriate constraint on the path integral.

In what follows, we shall adopt the second approach. In the massless limit, the zero mode on the fuzzy sphere is the constant mode $\phi_0 I$, associated with the trace of the matrix field, $\phi_0 = \mathrm{Tr}\Phi$. This mode represents a global degree of freedom shared by all regions or, in our case, uniformly distributed over both halves of the system. Since it is completely delocalized, it cannot be naturally assigned to either subsystem under the bipartition \footnote{Technically, there is no way to decompose this mode into two parts, each belonging to one of the two halves of the sphere.}
. It is therefore physically reasonable to define the reduced density matrix in terms of the local fluctuations alone, with the global zero mode projected out . This allows us to isolate the genuinely spatially entangling degrees of freedom and avoid the contamination of the EE by the global collective fluctuation.

To this end, we reconsider the derivation given in \cite{Allouche:2018err,Allouche_2022} for the $m=0$ sector and impose the constraint $\mathrm{Tr}\Phi=0$  by inserting $\delta[\mathrm{Tr}\phi]$ into the path integral defining the reduced density matrix. In terms of the notation introduced in \eqref{cv1}, this is equivalent to imposing the constraint $ \sum_{i=1}^{N} Q_i^{(0)}=0$.   The resulting steps are straightforward: integrating over the delta function eliminates the zero mode and leads to the following intrinsic $n$-fold action for the nonzero modes in the $m=0$ sector:

\be\label{actionzeromode}
S^{(0)}=\int_{-\infty}^{\infty}   \frac{1}{2} [-\mathbb{Q}^T \frac{d}{d\tau}\mathbb{M}(\tau) \frac{d}{d\tau}{\mathbb{Q}} +\mathbb{Q}^T\tilde{\mathbb{V}}^{(0)}(\tau)\mathbb{Q}] d\tau
\ee

where
$$
\mathbb{M}(\tau)= \theta (-\tau) \mathbb{M}_-+ \theta (\tau) \mathbb{M}_+,~~~\mathbb{M}_-=\mathbb{I}_{n} \otimes M, ~~\mathbb{M}_+=\mathbb{P}_\pi^T\mathbb{M}_-\mathbb{P}_\pi 
$$

$$
M_{ij}=\delta_{ij}+1, ~~i,j=1,2,\cdots {N-1}
$$
and

$$
\tilde{\mathbb{V}}^{(0)}(\tau)= \theta (-\tau) \tilde{\mathbb{V}}^{(0)}_-+ \theta (\tau) \tilde{\mathbb{V}}^{(0)}_+,~~~\tilde{\mathbb{V}}^{(0)}_-=\mathbb{I}_{n} \otimes \tilde{V}^{0}, ~~\tilde{\mathbb{V}}^{(0)}_+=\mathbb{P}_\pi^T\tilde{\mathbb{V}}^{(0)}_-\mathbb{P}_\pi 
$$
$$
\tilde{V}^{(0)}_{ij}=V^{(0)}_{ij}+V^{(0)}_{NN}-V^{(0)}_{iN}-V^{(0)}_{Nj}~~i,j=1,2,\cdots {N-1}
$$
$\mathbb{P}_\pi$ is the permutation matrix associated with tracing out half of the remaining degrees of freedom corresponding to the diagonal elements of $\Phi$.\footnote{We have chosen to integrate out the element $\phi_{NN}$ in order to implement the constraint. However, any other diagonal element could be chosen instead. Our numerical calculations show that the resulting correction to the EE is independent of this choice.}

The above action, \eqref{actionzeromode}, leads to a Green's matrix function satisfying the following equation:

\begin{equation}\label{GEzeromode}
	\big[-\frac{d}{d\tau}\mathbb{M}(\tau) \frac{d}{d\tau} +\tilde{\mathbb{V}}^{(0)}(\tau) \big]\mathbb{\tilde{G}}^{(0)}(\tau,\tau')=\delta (\tau-\tau').
\end{equation}

The above differential equation differs from that considered in \cite{Allouche:2018err, Allouche_2022}. Nevertheless, the corresponding Green's function can be constructed using the solutions given in Eqs.~\eqref{GFS} and~\eqref{GFS2}. In the appendix~\ref{Appendix II}, we outline the essential steps for constructing the Green's function for this class of problems and provide its explicit form in terms of $\mathbb{M}$ and $\tilde{\mathbb{V}}^{(0)}$.\footnote{Note that both $\mathbb{M}$ and $\tilde{\mathbb{V}}^{(0)}$ are positive definite.}

Having constructed the Green's function for the $m=0$ sector, it is straightforward to determine how Eq.~\eqref{cotractiontnegative} is modified when the zero mode (more precisely, the would-be zero mode) is projected out. The corresponding zero-mode-free expectation value of $\mathbb{V}_{int}(\tau)$ for $\tau<0$ is obtained by making the following substitutions:

\begin{eqnarray}
	\sum_{i=1}^{N}\big(\mathbb{G}^{(0)}_{ii}\big)^2&\longrightarrow & \sum_{i=1}^{N-1}\big(\tilde{\mathbb{G}}^{(0)}_{ii}\big)^2+\sum_{i,j,k,l=1}^{N-1}\tilde{\mathbb{G}}^{(0)}_{ij} \tilde{\mathbb{G}}^{(0)}_{lk}.\nonumber\\
	\sum_{l=1}^{N-1}\sum_{i=1}^{N-l}\mathbb{G}_{l+i,l+i}^{(0)}\mathbb{G}^{(l)}_{ii} &\longrightarrow& \sum_{l=1}^{N-2}\sum_{i=1}^{N-l-1}\tilde{\mathbb{G}}_{l+i,l+i}^{(0)}\mathbb{G}^{(l)}_{ii}+ \sum_{i,j,l=1}^{N-1}\tilde{\mathbb{G}}_{i,j}^{(0)}\mathbb{G}^{(l)}_{N-l,N-l}.\nonumber\\
\sum_{m=1}^{N-1}\sum_{l=0}^{N-m-1}\sum_{i=1}^{N-m-l}\mathbb{G}_{ii}^{(l)}\mathbb{G}_{l+i,l+i}^{(m)}&\longrightarrow& \sum_{m=1}^{N-1}\sum_{l=1}^{N-m-1}\sum_{i=1}^{N-m-l}\mathbb{G}_{ii}^{(l)}\mathbb{G}_{l+i,l+i}^{(m)}+\sum_{m=1}^{N-1}\sum_{i=1}^{N-m}\tilde{\mathbb{G}}_{ii}^{(0)}\mathbb{G}_{ii}^{(m)}.\nonumber\\
\sum_{m=1}^{N-1}\sum_{l=0}^{m}\sum_{i=m+1}^{N}\mathbb{G}_{i-l,i-l}^{(l)}\mathbb{G}_{i-m,i-m}^{(m)} &\longrightarrow& \sum_{m=1}^{N-1}\sum_{l=1}^{m}\sum_{i=m+1}^{N}\mathbb{G}_{i-l,i-l}^{(l)}\mathbb{G}_{i-m,i-m}^{(m)}\nonumber\\
&+&\sum_{m=1}^{N-2}\sum_{i=m+1}^{N-1}\tilde{\mathbb{G}}_{i,i}^{(0)}\mathbb{G}_{i-m,i-m}^{(m)}.
+
\sum_{i,j,m=1}^{N-1}\tilde{\mathbb{G}}_{i,j}^{(0)}\mathbb{G}_{N-m,N-m}^{(m)}.\nonumber\\
\sum_{m=0}^{N-2}\sum_{l=1}^{N-m-1}\sum_{i=l+1}^{N-m}\mathbb{G}_{i,i-l}^{(m)}\mathbb{G}_{i-l,i-l+m}^{(l)} &\longrightarrow&  \sum_{m=1}^{N-2}\sum_{l=1}^{N-m-1}\sum_{i=l+1}^{N-m}\mathbb{G}_{i,i-l}^{(m)}\mathbb{G}_{i-l,i-l+m}^{(l)}\nonumber\\
&+&\sum_{l=1}^{N-2}\sum_{i=l+1}^{N-1}\tilde{\mathbb{G}}_{i,i-l}^{(0)}\mathbb{G}_{i-l,i-l}^{(l)}
+\sum_{i,l=1}^{N-1}\tilde{\mathbb{G}}_{i,N-l}^{(0)}\mathbb{G}_{N-l,N-l}^{(l)}.\nonumber
\end{eqnarray}

The remaining terms are unchanged, since they do not involve the $m=0$ sector. The integral over $\tau$ is again evaluated using the spectral decomposition of $\mathbb{W}^{(m)}$. We then define the following dimensionless rescaled quantity:

\be\label{fBzeromode}
\tilde{f}_n(\mu,N)=\frac{4!}{\l NR}\frac{\delta \ln \mathrm{Tr} \tilde{\rho}_A^n}{(n-1/n)},
\ee

and evaluate it numerically for different values of $n$ and $\mu$ as a function of $N$.

\begin{figure}[h]
	\centering
	\includegraphics[trim=0cm 8cm -1.5cm 8cm, clip, width=1.1\linewidth]{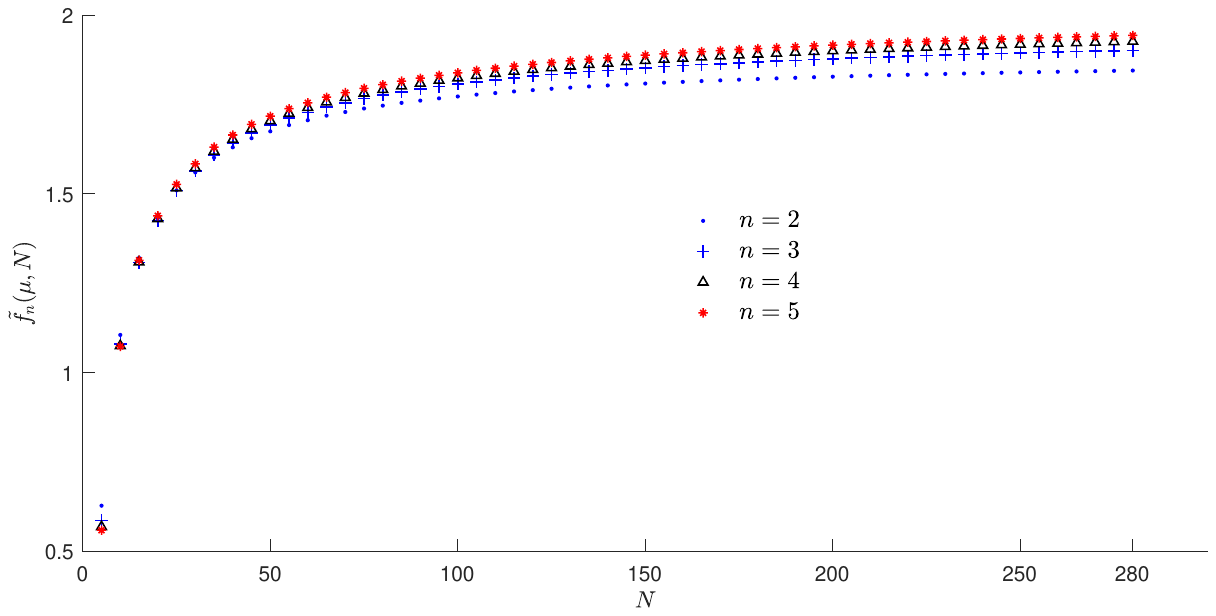}
	\caption{The zero-mode-projected rescaled function $\tilde{f}_n(\mu,N)$ plotted as a function of $N$ for $\mu=0.001$ and different values of $n$.}
	\label{FigSNZM}
\end{figure}

 In Figue~\ref{FigSNZM}, we plot $\tilde{f}_n(\mu,N)$ for $n=2,3,4,5$ and $\mu=0.001$.The numerical results demonstrate clear convergence toward a constant value as $N$ increases. The curves corresponding to different values of $n$ differ only slightly, with these differences becoming independent of $N$ for larger values of $N$.
 The dependence on the mass is likewise very weak, as confirmed by evaluating $\tilde{f}_n(\mu,N)$ for several values of $\mu$. For instance we have:
$$
\tilde{f}_n(\mu=0.0001,N)-\tilde{f}_n(\mu=0.001,N)\sim 10^{-7}, ~~\text{for}~N=200.
$$

Of course, this remains true as long as the mass is not too large. A sufficiently large mass renders the off-diagonal terms in the potentials $V^{(m)}$ responsible for the entanglement negligible compared with the diagonal terms, thereby suppressing the entanglement.

We now turn our attention to the $n$-dependence of $\tilde{f}_n(\mu,N)$. In the large-$N$ limit, the difference between $\tilde{f}_n(\mu,N)$ for different values of $n$ becomes independent of $N$. This property allows us to infer the scaling form of the leading contribution to EE. Choosing $n=2$ as a reference point, we decompose $\tilde{f}_n$ as
\begin{equation}
	\tilde{f}_n(\mu,N) = \tilde{f}_2 + c_n, \qquad c_2 = 0, \qquad \tilde{f}_2 \approx 1.84 \quad (N \gg 1).
\end{equation}

Numerically, we find that $\Delta c_n=c_{n+1}-c_n\sim0.05$ for $n=2$, which is the largest value of $\Delta c_n$, for $n\ge 2$. As $n$ increases, $\Delta c_n$ monotonically vanishes ($\Delta c_n \to 0$ as $n \to \infty$). This behavior is consistent with approaching the min-entropy limit ($n \to \infty$), where the state is dominated by the largest eigenvalue of the reduced density matrix.

Consequently, the first-order correction to  R\'enyi entropy in the large-$N$ regime takes the  form
\begin{equation}\label{FOCZmode}
	\delta \ln \operatorname{Tr} \tilde{\rho}_A^n = \frac{\lambda}{4!}\left(n-\frac{1}{n}\right) NR\left(\tilde{f}_2+c_n\right) + \dots,
\end{equation}
where the ellipsis represents subleading terms depending on both $n$ and the mass parameter $\mu$.

Taking the derivative at $n=1$ yields the leading correction to the entanglement entropy:
\begin{equation}\label{En+corr1}
\delta S_{ent}^{(1)}=	-\left. \frac{\partial}{\partial n} \delta \ln \operatorname{Tr} \tilde{\rho}_A^n \right|_{n=1} = -\alpha\lambda NR + \dots,
\end{equation}
and hence
\begin{equation}\label{En+corr}
	S_{\mathrm{ent}} = 0.39 N - \alpha\lambda NR + \dots,
\end{equation}
where
\begin{equation}
	\alpha = \frac{1}{12}\left(\tilde{f}_2+c_1\right).
\end{equation}

The constant $c_1 \equiv c_n|_{n=1}$ cannot be extracted directly from numerical data at integer $n\ge 2$. However, assuming a smooth analytic continuation of $c_n$ down to $n=1$ based on its observed monotonicity, a spline extrapolation yields   the estimate   
$	c_1 \approx -0.1 $, corresponding to $\alpha \approx 0.14$\footnote{In fact, the $N$-independence of $\Delta c_n$ is sufficient to establish the exact $N$-scaling of the correction; determining $c_1$ is only necessary for fixing its numerical coefficient.}.

Unlike the free entanglement entropy, the first-order interaction correction for a bipartition of the fuzzy sphere is no longer directly proportional to the number of boundary degrees of freedom. Consequently, it cannot be interpreted simply as a quantity proportional to the circumference of the entangling boundary.

Although the quartic interaction  leads to a violation of the area law, the degree of UV divergence in the commutative continuum limit remains the same as in the free Gaussian theory. Indeed, the effective short-distance (UV) cutoff on the fuzzy sphere is approximately given by $\epsilon=R/N$. We may therefore write
 \be\label{En+corr2}
 S_{ent}=0.39 \frac{R}{\epsilon}-\alpha\lambda \frac{R^2}{\epsilon}
 +\cdots ,
 \ee

The free contribution and the interaction correction exhibit  the same degree of UV divergence in the commutative continuum limit. However, their geometric scaling is fundamentally different. While the free contribution is proportional to the size of the entangling boundary, the interaction correction is proportional to the volume of the bulk region, i.e., to the area of the hemisphere. Thus, although both terms share the same degree of UV divergence, the interaction correction has an intrinsically extensive character, rather than the boundary scaling characteristic of the free theory. We discuss this extensive behavior and its origin in the Discussion section.

It is also worth noting that, although the factor $\lambda R$ appearing in the correction is essentially fixed by dimensional analysis, its dependence on $N$—and, equivalently, the degree of UV divergence of the interaction correction—cannot be determined \emph{a priori}.

\subsection{Interaction Correction on the Fuzzy Disc}

Let us now consider the first-order correction in the fuzzy disc model. As mentioned earlier, there are two possible bipartitions to consider. The first is algebraically similar to the one considered above for the fuzzy sphere. However, in the oscillator basis, the subset of degrees of freedom being traced out has no clear or natural geometrical interpretation. We therefore do not consider the interaction correction for this bipartition.
The second, and more interesting, case is obtained by tracing out a sub-disc of smaller radius $r$. In the oscillator basis, the degrees of freedom are naturally localized in the radial direction, with each oscillator state associated with a definite value of $r^2=x^2+y^2$. Consequently, a sub-disc provides a well-defined geometrical region to be traced out, giving rise to a natural geometrical bipartition and the corresponding EE.

The expression for the first-order correction in the sub-disc case differs slightly from that obtained for the fuzzy sphere, although the overall formalism and the intermediate steps are essentially the same. We therefore do not repeat the derivation here and quote the final result in the appendix.

The quantity we evaluate numerically is defined as follows:

\be\label{fnmuNHD}
f_n(\mu,N)=\frac{4!}{\l r_q}\frac{\delta \ln\mathrm{Tr} \r_A^n}{ (n-1/n)}
\ee
where $r_q=\sqrt{q\theta}$ is the radius of the sub-disc, and $q$ specifies the corresponding $q\times q$ submatrix \eqref{subdiscmatrix} representing the degrees of freedom to be traced out.
\begin{figure}[htbp]
	\centering
	\includegraphics[trim=0cm 6cm 0cm 6cm, clip, width=1.1\linewidth]{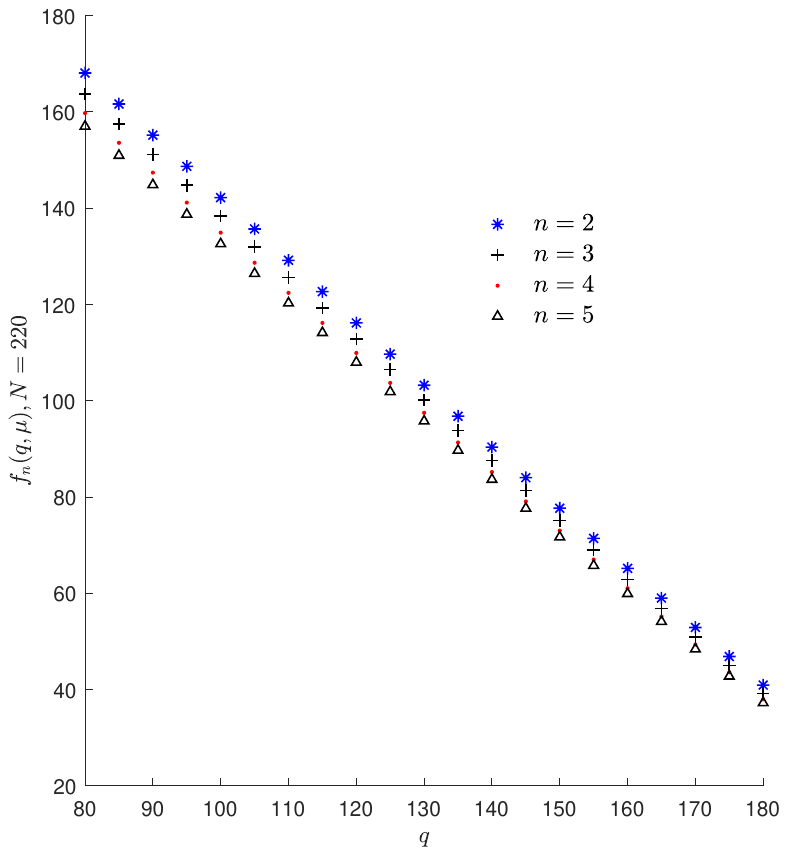}
\caption{The rescaled dimensionless function ${f}_n(q,\mu,N)$ plotted as a function of $q$ for $\mu=0.001$, different values of $n$, and $N=220$.}
	\label{FigSD}
\end{figure}

Figure~\ref{FigSD} shows $f_n(\mu,q,N)$ for $n=2,3,4,5$, with $\mu=0.001$ and $N=220$, as a function of $q$ over the range $80\leq q\leq 180$. It is readily apparent that the three curves corresponding to different values of $n$ are well fitted by straight lines with nearly identical slopes. As $n$ increases, the curves shift slightly downward, with the magnitude of the shift decreasing as $n$ increases.

Again, our numerical calculations show that the dependence on the mass is very weak and can be neglected at the level of the subleading contributions. It is worth noting that, in the fuzzy disc case, there is no need to project out any zero mode, since such a collective zero mode is absent. Although the identity operator commutes with the Laplacian, the presence of the boundary prevents the Laplacian from developing a zero-mode in the massless limit. Consequently, the interaction correction is no longer dominated by the mass term and exhibits only weak IR sensitivity.

As for the $n$-dependence of $f_n$, the same observations made in the fuzzy sphere case apply here. In particular, the difference
$\Delta f_n=f_{n+1}-f_n$
becomes approximately independent of both $N$ and $q$ in the regime $N\gg q\gg1$, and tends to zero as $n\rightarrow\infty$. More precisely, for a large sub-disc embedded in a much larger disc, our numerical results are well described by the following set of straight lines:
\be\label{ffit1}
f_n(N,q,\mu)=b(N-q)+c_n,
\ee
where $b$ is a numerical constant, found to be approximately $b\simeq1.27$, while $
c_2\simeq-11,
c_3\simeq-15,
c_4\simeq-18.2$ and
$c_5\simeq-19.5$.

It is important to stress that Eq.~\eqref{ffit1} cannot hold over the entire range of \(q\). In particular, as \(q\) approaches \(N\), the correction must deviate from the above linear behavior and eventually vanish in the limiting case \(q=N\). Although our numerical analysis shows that the linear behavior remains a good approximation up to values of \(q\) relatively close to \(N\), deviations eventually become significant as \(q\to N\).

As \(q\) approaches the other limiting case, \(q=0\), \(f_n\) itself need not vanish, since the correction is obtained by multiplying it by \(\sqrt{q}\). Nevertheless, \(f_n\) also shows a departure from the linear behavior in this regime. For \(N=220\), our numerical results show a slight departure from linearity already around \(q\simeq35\), followed by a more pronounced deviation and a decrease of \(f_n\) as \(q\) is reduced below \(q\simeq15\).

Taken together, these results imply that, for \(N\gg q\gg1\), the leading contribution to the first-order correction to the Rényi entropy is given by


\be\label{correSD}
\delta \ln \mathrm{Tr}{\rho}_A^n=  \frac{\lambda \sqrt{\theta}}{4!} \left(n-\frac{1}{n}\right)\sqrt{q}\big[b(N-q)+c_n\big]+\cdots.
\ee

\begin{figure}[htbp]
	\centering
	\includegraphics[trim=0cm 8cm 0cm 8cm, clip, width=1.1\linewidth]{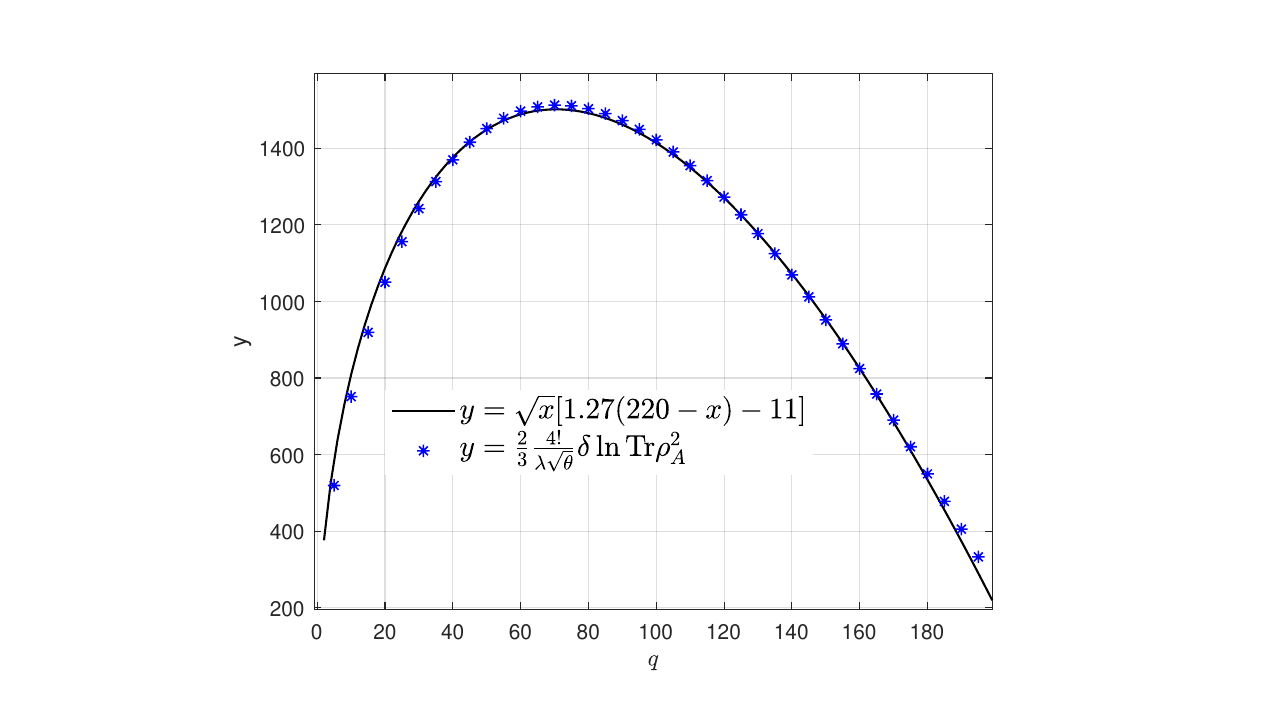}
	\caption{Exact values of the correction for $n=2$ compared with the curve representing the leading contribution given by Eq.~\eqref{correSD}, for $N=220$ and $5\le q\le 200$.}
	
	\label{FigFit}
\end{figure}

Although the above leading contribution was derived using the linear approximation valid in the regime \(N\gg q\gg1\), it turns out to provide a very good approximation to the full interaction correction and remains accurate even close to the limiting cases \(q=N\) and \(q=0\). Figure~\ref{FigFit}  shows the exact values of $\delta\ln\mathrm{Tr}\rho_A^n$, together with the curve obtained from the leading term of Eq.~\eqref{correSD} for $n=2$.

The leading contribution to the entanglement entropy for $N\gg q\gg 1$, including the first-order interaction correction, can then be written as
\be\label{EnHD+corr2}
S_{ent}
=0.46 q-\b {\lambda} r_q(N-q)+\cdots,~~\b\simeq 0.2.
\ee

Several remarks are in order.

Unlike the free (Gaussian) entanglement entropy, which is determined solely by the DF associated with the entangling surface and is therefore directly proportional to $q$, the interaction correction depends on the geometry of the entire disc, involving both the interior and exterior regions separated by the entangling surface.

To make this distinction more explicit, recall that only the sectors with $|m|<q$ contribute to the free EE. The sectors with $|m|\geq q$, on the other hand, are not at the free level entangled with the DF belonging to the sub-disc and therefore have trivial Green's functions\footnote{ $\mathbb{G}_{--}=\mathbb{G}_{++}=\frac{1}{2\mathbb{W}_-} e^{-\mathbb{W}_-|\t-\t'|} ,  ~(\mathbb{W}_-=\mathbb{W}_+)$.}. Consequently, for large $N$ and $q$, the total number of entangled DF at the free level, including those in both the interior and exterior regions, is $d_E=2qN-q^2$. 

In terms of $N$ and $d_E$, the leading first-order correction can therefore be written as

\be\label{Correction21}
\delta S^{(1)}_{ent} =-2\b{\lambda} L_{\partial A} \mathcal{G}_{\text{bulk}},
\ee

where $  L_{\partial A}=r_q, ~~\mathcal{G}_{\text{bulk}}= \sqrt{N^2-d_E}$

Expressing $\delta S^{(1)}_{ent}$ in terms of $r_q$, $N$, and $d_E$ makes two interesting features of the first-order correction manifest. First, the entanglement entropy satisfies the symmetry 
$$
S_{ent}(\rho_A)= S_{ent}(\rho_{\bar{A}})
$$ 

which must likewise be respected by the interaction correction. Equation~\eqref{Correction21} makes this symmetry explicit: $L_{\partial A}$ is simply the length of the entangling boundary, while the factor $\mathcal{G}_{\text{bulk}}$ depends only on the total number of DF of the fuzzy disc and on the total number of entangled DF shared between the sub-disc and its complement. Consequently, $\mathcal{G}_{\text{bulk}}$ is invariant under the exchange $A\longleftrightarrow\bar{A} $.

Qualitatively, the first-order correction may therefore be interpreted as a boundary factor multiplied by a genuinely bulk-dependent factor. This reflects the fact that, unlike the free EE, the interaction correction is sensitive not only to the DF in the immediate vicinity of the entangling boundary, but also to the DF distributed throughout the entire fuzzy disc. Formally, we may write
\be\label{CorreGeometry}
\delta S^{(1)}_{ent}
=\lambda \times \text{Boundary Contribution}
\times \text{Bulk Geometry Factor}.
\ee

This interpretation is further supported by the limiting cases. Although the leading contribution was derived under the assumption that \(q\) is sufficiently far from the two limiting cases, Eq.~\eqref{Correction21} correctly reproduces both of them. The bulk factor \(\mathcal{G}_{\text{bulk}}\) vanishes when \(A=D_N\), corresponding to tracing out the entire disc; in this case, there is no complementary subsystem, and the EE must vanish. Conversely, when \(A=\emptyset\), the boundary factor vanishes, and the interaction correction vanishes accordingly.


Before moving on, let us take a step back and reconsider the result obtained for the fuzzy sphere. Here, the fact that we traced out one half of the sphere, leading to the result given in Eq.~\eqref{En+corr1}, tends to obscure an interpretation analogous to that developed for the fuzzy disc. To make this analogy more transparent, consider instead tracing out a polar cap, as illustrated in Fig.~\ref{figFvC}. In this case, one would expect the first-order correction to take the form

$$
\delta S^{(1)}_{\rm ent}\sim \lambda\, q R,
$$

where $q$ denotes the number of boundary degrees of freedom associated with the polar cap. The value of $q$ is determined by the polar angle $\theta$ defining the size of the cap. Thus, as in the fuzzy disc, the correction can be viewed as consisting of a boundary factor, represented by $q$, multiplied by a bulk geometric factor proportional to $R\).

Another important point in this context concerns the contribution of the unentangled DF to $\delta S^{(1)}_{ent}$. As noted above, in the free theory, all DF belonging to sectors with $|m|\ge q$ make no contribution to $S^{(0)}_{ent}$. A natural question is whether these unentangled DF can nevertheless contribute to the first-order interaction correction.

Although the analytical expression for $\delta\ln\mathrm{Tr}\rho_A^n$ contains explicit terms from sectors with $|m|\ge q$, its form does not readily reveal how significant these are relative to the entangled sectors. However, this can be addressed by examining the individual contributions numerically.
To this end, we decompose $f_n(N,q)$ into two parts, corresponding respectively to the contributions from the entangled and unentangled sectors:
\be\label{f_E}
f_n(N,q)=f_E+f_{NE}.
\ee

 where $f_E$ denotes the contribution arising exclusively from the entangled sectors, $|m|\le q$, while $f_{NE}$ contains the mixed contributions involving the unentangled sectors.\footnote{The unentangled sectors do not contribute independently. Owing to their trivial Green's functions, their contributions arise only indirectly through their coupling to the entangled sectors via the interaction.} Several numerical calculations show that the contribution of each term to the total correction is approximately proportional to the number of DF in the corresponding sectors. For example, for $N=160$ and $q=20$, the number of entangled DF is $d_{E}=23\%N^2$, while $f_E$ accounts for approximately $42\%\ $ of the total correction to the EE. For $q=30$, we find $d_E=33\%N^2$, with $f_E$ accounting for approximately $53\%$ of the correction. Finally, for $q=70$, $d_E=68\%N^2$, while $f_E$ accounts for approximately $85\%$ of the total correction, ..etc.
 
 These results provide strong evidence that the interaction correction is governed by the geometry of the entire disc—or, equivalently, by its bulk DF—rather than being localized to any particular region or subset of DF. We return to this observation and discuss its implications in more detail in the Discussion section.

A second noteworthy feature of the above result concerns its behavior in different limits. For the fuzzy disc, there are two distinct and physically interesting limits that are worth considering.

We first consider the commutative, or continuum, limit, in which the fuzzy disc approaches the classical disc. In this limit, $\theta \rightarrow 0$, while $N,q \rightarrow \infty$, with $N\theta = R^2$ and $q\theta = r^2$ held fixed. The largest eigenvalue of the fuzzy Laplacian scales as $\frac{N^2}{R^2}$, corresponding to a UV momentum cutoff $\frac{N}{R}$, . Introducing $\xi=\frac{r}{R}$, the ratio of the radius of the sub-disc to that of the full disc, the free EE can then be written as
\be\label{FreeEEcontinum1}
S_{eng}(\lambda =0) \sim \frac{2\pi r}{\epsilon},~~~ ~~~ \epsilon\sim \frac{R}{N}
\ee

By the same considerations, the leading term of the first-order correction to the EE in the continuum limit takes the form
\be\label{EEcorDisc3}
\delta S_{ent} \sim -
\lambda \frac{r}{\epsilon}(1-\xi^2)R.
\ee

Again, the interaction correction exhibits the same degree of UV divergence as the free EE. Unlike the free contribution, however, it does not obey a pure area law. Instead, it contains an additional factor that depends on the geometry of the entire disc. This behavior is fully consistent with the geometric interpretation of the interaction correction developed above in the noncommutative setting.

It is worth emphasizing, however, that this result is in sharp contrast with what one would expect for an interacting theory formulated directly in ordinary commutative space.

For instance, Ref.~\cite{Hertzberg:2012mn} showed that, in $(d+1)$-dimensional flat spacetime, the first-order correction to the entanglement entropy (EE) in a $\lambda\phi^4$ theory, obtained by tracing out a half-space, contains no volume divergence. Moreover, it was found that, at least to first order in $\lambda$, the full EE retains the same functional form as in the free theory, with the bare mass replaced by the renormalized mass.

Our result, Eq.~\eqref{EEcorDisc3}, does not, however, reproduce this continuum behavior in the commutative limit. This discrepancy indicates that the operations of computing the interaction correction to the EE on a fuzzy space and subsequently taking the commutative continuum limit do not, in general, commute with the operation of computing the corresponding correction directly in the continuum theory with a UV regulator. In other words, the continuum limit of the fuzzy-space result need not coincide with the result obtained from the continuum theory itself.

Let us now turn to a more interesting limit, namely the Moyal plane limit. We fix $q$ at a sufficiently large value for our numerical scaling relation to remain valid and consider the limit 

$$
N \rightarrow \infty,~~~~\theta ~\text{fixed}
$$ 

In this limit, the radius of the fuzzy disc grows without bound, and the disc approaches the Moyal plane. The entangling sub-disc, whose radius $r_q=\sqrt{q\theta}$ remains fixed, thus becomes a finite region embedded in an increasingly large noncommutative disc. The EE, including the first-order interaction correction, then takes the form

\be\label{EEcorDisc4}
 S_{ent} =0.46 q-\b{\lambda} r_qN=0.46q -\b{\lambda} r_q\frac{R^2}{\theta}
\ee

We see that the free (zeroth-order) EE remains finite in the Moyal plane limit. This is consistent with the result established in Ref.~\cite{Dou:2009cw}, namely that, at the free level, the EE is localized in the vicinity of the entangling boundary. By contrast, the first-order interaction correction grows with the total area of the disc and therefore diverges in the Moyal plane limit.

This result is somewhat surprising, as it implies that the EE across the entangling boundary is sensitive to degrees of freedom (DF) located arbitrarily far from the boundary, rendering the interaction correction intrinsically nonlocal. This is consistent with the observations made above, in connection with Eq.~\eqref{f_E}, which show that DF far from the entangling surface can make a significant contribution to the correction.

This pronounced infrared nonlocality is reminiscent of the well-known UV/IR mixing encountered in interacting noncommutative field theories. However, the infrared divergence found here is neither caused by a vanishing mass nor does it, at first sight, appear to originate from the same mechanism responsible for UV/IR mixing in noncommutative quantum field theories, whereby ultraviolet loop momenta generate singular behavior in the infrared regime \cite{Minwalla_2000}. Rather, as far as our numerical analysis indicates, the divergence is an \emph{infrared-extensive} effect: it arises from the accumulation of an approximately constant contribution over successive radial shells, including annuli located arbitrarily far from the entangling boundary.

Evidence supporting this interpretation, together with a possible connection between this infrared-extensive behavior and the UV/IR mixing phenomenon, will be discussed in detail below.

\section{Discussion}
In this paper, we have numerically investigated the first-order correction to the Rényi entropy induced by a quartic interaction in a scalar field theory on $2+1$-dimensional fuzzy spaces. We considered two fuzzy regularizations: the fuzzy sphere and the fuzzy disc. For the fuzzy sphere, the Rényi entropy was computed by tracing out one hemisphere, while for the fuzzy disc, we considered a radial bipartition in which a sub-disc of radius $r$ is traced out from a fuzzy disc of radius $R$. In both models, our numerical results allowed us to extract the analytic form of the leading contribution to the first-order correction to the Rényi entropy, and consequently to the EE.

  In the fuzzy sphere case, we found that the interaction correction explicitly violates the area law. Rather than being determined solely by the entangling surface (circle), it receives comparable contributions from the bulk DF in the interior and exterior regions, leading to a correction that scales with the volume (area) of the entire sphere, while exhibiting the same degree of UV divergence in the continuum limit as the free EE. The mechanism underlying this apparently extensive behavior, namely the additional bulk contribution proportional to $R$, however, remains to be understood.
  
  This behavior is in sharp contrast with that of the free theory. It was shown in Ref.~\cite{Dou:2009cw} that, at the free level, the EE is almost entirely accounted for by the entanglement between the inner and outer DF located arbitrarily close to the entangling circle  for large $N$. More precisely, this was established by computing the EE while retaining only the DF lying within a distance of $d$ lattice spacings from the entangling boundary, i.e., $d$ DF on each side of the boundary. Operationally, this was achieved by setting to zero all off-diagonal entries of the potential matrix except those involving the selected $2d+1$ DF \footnote{Some angular sectors exhaust their maximum contribution before the number $2d+1$ is reached.} . Equivalently, we retain all DF except those residing in the upper and lower polar caps of the sphere, with opening angle $\theta$, and progressively increase $\theta$; see Fig.~\ref{figFvC} for an illustration. The resulting EE was found to reproduce the full EE with remarkable accuracy, demonstrating that, at the free level, entanglement is essentially localized in the vicinity of the entangling circle .

We now turn to the apparent extensivity of the interaction correction. In particular, we seek to determine whether the breakdown of the area law is merely the result of an enhancement of the near-boundary contributions already present in the free theory, or whether it instead reflects an intrinsically IR-nonlocal contribution to the entanglement.

To this end, we perform the same analysis described above, but now for the interaction correction. We compute a truncated correction, $\delta \ln \mathrm{Tr}\rho_A^n(d)$, by retaining only the DF located within a distance $d$ (in lattice units) of the entangling boundary. We then progressively increase $d$, thereby incorporating DF at increasingly larger distances from the boundary, until all DF on the sphere are included. For each value of $d$, we compute the ratio of the truncated correction to the full interaction correction,
 \[
 \mathcal{R}_d =
 \frac{\delta \ln \mathrm{Tr}{\rho}_A^n(d)}
 {\delta \ln \mathrm{Tr}{\rho}_A^n},
 \]
 which measures the fraction of the full correction accounted for by the DF included within a distance $d$ of the entangling boundary.

 \begin{figure}[htbp]
 	\centering
 	\includegraphics[trim=0cm 8cm 0cm 8cm, clip, width=1.1\linewidth]{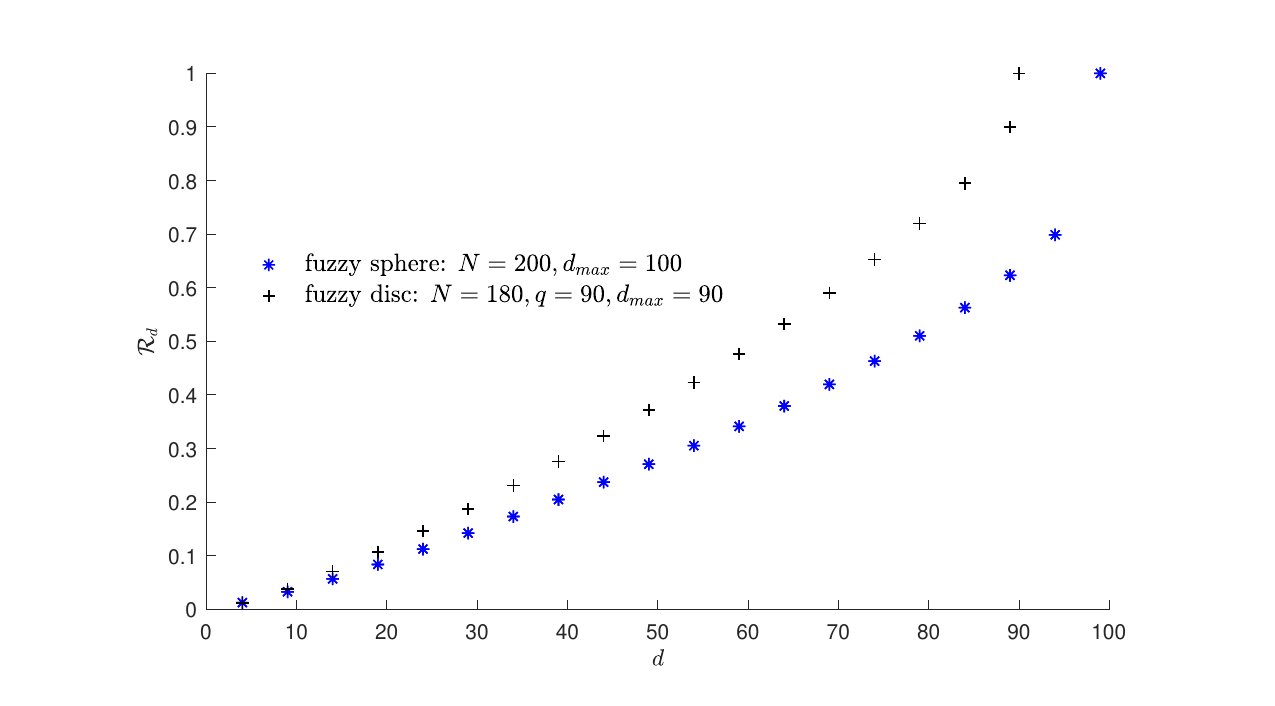}
 	\caption{Fraction of the total first-order correction as a function of the number of incorporated DF for the fuzzy sphere and fuzzy disc. \(d_{\max}\) denotes the value of \(d\) for which all DF are included.}
 	
 	\label{FigNB}
 \end{figure}

 The results shown in Fig.~\ref{FigNB} clearly demonstrate that the near-boundary DF alone do not account for any  substantial fraction of the full interaction correction. As increasingly distant DF are included, the truncated correction grows progressively, with the full correction being recovered only when essentially all DF on the sphere are included. More quantitatively, our numerical results indicate that $\mathcal{R}_d$ is approximately proportional to the fraction of DF retained within a distance $d$ of the entangling boundary, namely $n_{DF}/N^2$. The proportionality coefficient depends only weakly on $d$ and approaches an approximately constant value for sufficiently large $d$. For example, $\mathcal{R}_9=0.032$ while $n_{DF}/N^2=0.087$; $\mathcal{R}_{64}=0.37$ while $n_{DF}/N^2=0.53$; and $\mathcal{R}_{94}=0.69$ while $n_{DF}/N^2=0.71$. In particular, for $d=99$,  when only one DF remains to be included from each of the inner and outer regions, we recover only $99\%$ of the total correction.
 
 Thus, unlike the free EE, for which the dominant contribution is concentrated near the entangling boundary, the interaction correction receives appreciable contributions from DF throughout the bulk of the fuzzy sphere.

Let us now address the same question in the fuzzy disc and examine how the extensive character of the interaction correction is built up in this case. We first recall that, at the free level, it was too shown in Ref.~\cite{Dou:2009cw} that the EE is essentially accounted for by the DF in the vicinity of the entangling boundary.

We follow the same procedure described above by computing a truncated interaction correction, $\delta \ln \mathrm{Tr}\rho_A^n(d)$, obtained by retaining only the DF lying within an annulus centered on the entangling boundary, with outer radius $r_{q+d}$ and inner radius $r_{q-d}$. We progressively increase the width of the annulus by increasing $d$, thereby incorporating DF at increasingly larger distances from the entangling boundary, until all DF in the fuzzy disc are included. As in the fuzzy sphere case, for each value of $d$ we compute the ratio $\mathcal{R}_d$ of the truncated correction to the full interaction correction.

Figure~\ref{FigNB} shows a behavior remarkably similar to that observed in the fuzzy sphere case. The near-boundary approximation fails to capture a substantial fraction of the full interaction correction. As in the fuzzy sphere, we find that $\mathcal{R}_d$ is approximately proportional to the fraction of DF included within a distance $d$ of the entangling boundary, namely $n_{DF}/N^2$. Moreover, the full correction is not recovered, even approximately, unless essentially all DF on the fuzzy disc are included.

These results provide strong evidence that, in both the fuzzy sphere and the fuzzy disc, the extensive character of the interaction correction arises from the cumulative contributions of DF distributed throughout the entire fuzzy space. The correction is therefore not dominated by DF in the immediate vicinity of the entangling circle, but receives significant contributions from regions increasingly far from the entangling boundary. Indeed, the fact that the full correction is recovered only when essentially the entire set of DF is included demonstrates that this extensive behavior cannot be attributed merely to an amplification of local near-boundary contributions. Rather, it points to an intrinsically IR-nonlocal character of the interaction correction.

We conclude this paper by examining the infrared divergence that emerges in the Moyal plane limit and its possible connection with the well-known UV/IR mixing phenomenon, an issue that merits further discussion.

 If we assume that our result, Eq.~\eqref{EEcorDisc4}, can be extrapolated to the Moyal plane limit, then an IR divergence proportional to the volume (area) of the infinite space appears to be unavoidable.

Moreover, in both models, naive perturbation theory eventually breaks down as $N$ becomes sufficiently large. At fixed $N$, however, one can maintain a controlled perturbative expansion by requiring
$$
\lambda_{\mathrm{eff}}=\lambda\sqrt{\theta}N\ll1.
$$
 Equivalently, the issue can be understood as a non-uniformity of the perturbative expansion in $N$: there is no fixed $\lambda>0$ for which the first-order correction remains subdominant to $S^{(0)}_{ent}$ as $N$ is taken arbitrarily large. Thus, at fixed $\lambda>0$, taking the limit $N\rightarrow\infty$ eventually invalidates the perturbative expansion. This observation is particularly relevant for the Moyal plane limit, where $N\rightarrow\infty$ at fixed $\theta$.

This naturally raises  important questions. Is the apparently extensive IR divergence of the first-order correction a genuine pathology of the interacting theory when formulated with fuzzy-space regularization, or does it merely signal the breakdown of naive perturbation theory and the need for a non-perturbative treatment of the interaction contribution to the EE? In particular, determining whether the IR divergence persists after a suitable resummation or non-perturbative treatment is essential for understanding its possible relation to the UV/IR mixing characteristic of noncommutative field theories.

Let us emphasize that an entropy, including the entanglement entropy, scaling extensively with the volume of the whole space is not, in itself, problematic. What makes the present volume scaling potentially problematic is that it appears as a correction to a leading contribution that obeys an area law. In other words, the issue is not the extensive behavior per se, but the fact that an IR-extensive contribution emerges as a subleading interaction correction to an otherwise UV-dominated area-law term.

Another, perhaps more important, question concerns the origin of the IR divergence and its possible connection to the UV/IR mixing observed in non-commutative field theories . Let us recall that, in non-commutative field theories, IR divergences is generated by UV loop momenta, leading to singular behavior at small external momenta. Our IR divergence certainly resembles UV/IR mixing, in the sense that the one-loop (may be two loops) correction develops sensitivity to arbitrarily large distances. However, based on our present calculation alone, it is difficult to conclude that the underlying mechanism is the same.

To establish a genuine UV/IR mixing mechanism, one would ideally need to show that the $R^2$ dependence in Eq.~\eqref{EEcorDisc4} originates predominantly from the high spectral modes running in the loop. Although, as argued above and supported by our numerical results, the extensive contribution arises from summing an approximately constant contribution over all radial shells, it remains possible that the UV modes contribute substantially to an effect that ultimately grows as $R^2$.

Let us also note that our approach, together with the field decomposition adopted for this particular problem, does not allow, at least in its present form, for a straightforward diagrammatic representation of the first-order correction. Nor does it provide a direct analytical way of identifying the precise mechanism responsible for the IR divergence. Nevertheless, we can still make some heuristic observations.

In view of the explicit expression for the interaction given in the appendix, we can formally write the first-order correction in the schematic form
\be
\delta S_{ent}^{(1)}= \sum_{m,l,(a,b,c,d)}
F(\lambda_a^{(m)},\lambda_b^{(m)};\lambda_c^{(l)},\lambda_d^{(l)}).
\ee

Here, ${\lambda^{(m)}}$ denote the eigenvalues of $W^{(m)}$, namely the square roots of the eigenvalues of the potentials $V^{(m)}$, or equivalently the square roots of the eigenvalues of $\Delta_N+\mu^2$ in the corresponding angular momentum sector.

Although the sums over eigenvalues belonging to different angular momentum sectors are difficult to analyze analytically, the structure of the spectrum provides some indications that the UV modes may play an important role in generating the extensive scaling of the first-order correction.

For example, it can be shown that, for \(N\gg1\) and for angular-momentum sectors satisfying \(|m|=\mathcal{O}(1)\), the maximum eigenvalue is of order

$$
\l_{max}\sim\sqrt{ \frac{N}{\theta}}
$$,
which sets the UV scale, while the lowest eigenvalue is of order
$$
\l_{min}\sim\frac{\mu}{\sqrt{\theta}}
$$
 which sets the IR scale. Thus, in the low-angular-momentum sectors, the spectrum spans a  large range between the IR and UV scales.

For sectors with large $|m|$, but still satisfying $|m|\ll N$, the eigenvalues span a window whose IR cutoff is of order 
$\frac{m}{\sqrt{N\theta}} $ 
and UV cutoff remains of order
$ \sqrt{ N/\theta}$. 
 As $|m|$ approaches $N$, however, the situation changes qualitatively: for sectors with $|m|=\mathcal{O}(N)$,  one easily finds

 $$
 \l_{min}\sim \l_{max}\sim \sqrt{ N/\theta}
 $$
 
 so that these sectors are essentially composed entirely of UV-scale eigenmodes.

The numerical observation that the first-order correction exhibits little sensitivity to the mass provides an important clue in this context. It suggests that the deep IR part of the spectrum does not make a substantial contribution to the correction. In particular, the bulk DF belonging to the low-$|m|$ sectors that contribute to the correction are not predominantly associated with IR modes. As $|m|$ increases, an increasingly large fraction of the corresponding spectrum is pushed toward the UV, with the sectors satisfying $|m|\sim N$ becoming essentially composed entirely of UV modes. Nevertheless, as we have seen above, these sectors continue to contribute to the extensive sum that gives rise to the interaction correction.

Taken together, these observations do not provide a conclusive demonstration of a UV/IR mixing mechanism, but they do provide suggestive evidence that the extensive IR divergence observed in the Moyal plane limit may have a UV origin. In particular, the fact that increasingly UV-dominated angular momentum sectors continue to contribute to an IR-extensive correction raises the possibility that the divergence is a manifestation of UV/IR mixing in the fuzzy-disc regularization. Establishing this connection, however, would require a more direct analysis of the spectral contributions to the first-order correction.


\acknowledgments

The authors would like to thank B.H.Dou for his valuable assistance in optimizing the numerical codes used in the numerical computations presented in this paper.

\appendix 
 \section{ Explicit Expression for the First-Order Correction}\label{Appendix I}

 In this appendix, we provide the explicit form of the terms contributing to Eq.~\eqref{cotractiontnegative}, after performing the integration over $\tau$.
 
 The integration over $\tau$  can be carried out using the spectral decomposition of  $W^{(l)}$.
 In what follows, ${\lambda^{(l)}_a}$
  denote the eigenvalues of $W^{(l)}$, ${P^{(l)}_a}$   the corresponding projectors, and $T^{(m)}_{<a,b>}$ is $(N-m)\times (N-m)$ matrix defined as
  $$
  T^{(m)}_{<a,b>}=P_a^{(m)}M^{(m)}P_b^{(m)},
 $$ 
   where $M^{(m) }$ is the $(N-m)\times (N-m)$ matrix obtained as the first block of the block-circulant matrix $(\mathbb{W}^{(m)}_--\mathbb{W}^{(m)}_+)(\mathbb{W}^{(m)}_-+\mathbb{W}^{(m)}_+)^{-1}$.

{\small
\begin{eqnarray}
 	&& \int_{-\infty}^{0}d\tau\Big( \big(\mathbb{G}^{(0)}_{ii}\big)^2-\big(G^{(0)}_{ii}\big)^2\Big)=\frac{1}{4}\Big[ 
 	2\sum_{a,b=1}^{N}\frac{(W^{(0)})_{ii}^{-1}.\Big(T_{<a,b>}^{(0)}\Big)_{ii}}{\lambda_a^{(0)}(\lambda^{(0)}_a+\lambda^{(0)}_b)}+\quad\quad\nonumber\\
 	&&+
 	\sum_{a,b=1}^{N}\sum_{c,d=1}^{N}\frac{\Big(T_{<a,b>}^{(0)}\Big)_{ii}.\Big(T_{<c,d>}^{(0)}\Big)_{ii}}{\lambda_{a}^{(0)}\lambda_{c}^{(0)}(\lambda_{a}^{(0)}+\lambda_{b}^{(0)}+\lambda_{c}^{(0)}+\lambda_{d}^{(0)})}\Big].\nonumber\\
 &&\int_{-\infty}^{0}d\tau\Big(\mathbb{G}^{(l)}_{ii}\mathbb{G}^{(m)}_{l+i,l+i}-G^{(l)}_{ii}G^{(m)}_{l+i,l+i}\Big)=
 	\frac{1}{4}\Big[ 
 	\sum_{a,b=1}^{N-m}\frac{(W^{(l)})_{ii}^{-1}.\Big( T^{(m)}_{<a,b>}\Big)_{l+i,l+i}}{\lambda_a^{(m)}(\lambda^{(m)}_a+\lambda^{(m)}_b)}+\nonumber\\
&& \sum_{a,b=1}^{N-l}\frac{({W}^{(m)})_{l+i,l+i}^{-1}.\Big( T^{(l)}_{<a,b>}\Big)_{ii}}{\lambda_a^{(l)}(\lambda^{(l)}_a+\lambda^{(l)}_b)}+ 
 	\sum_{a,b=1}^{N-m}\sum_{c,d=1}^{N-l}\frac{\Big(T^{(m)}_{<a,b>}\Big)_{l+i,l+i}.\Big(T^{(l)}_{<c,d>}\Big)_{ii}}{\lambda_{a}^{(m)}\lambda_{c}^{(l)}(\lambda_{a}^{(m)}+\lambda_{b}^{(m)}+\lambda_{c}^{(l)}+\lambda_{d}^{(l)})}\Big]. \nonumber\\
 	&&\int_{-\infty}^{0}\Big(\mathbb{G}_{l+i,l+i}^{(0)}\mathbb{G}^{(l)}_{ii}-G_{l+i,l+i}^{(0)}G_{ii}^{(l)}\Big)=\frac{1}{4}\Big[\sum_{a,b=1}^{N-l}\frac{(W^{(0)})_{l+i,l+i}^{-1}.\Big(T^{(l)}_{<a,b>}\Big)_{ii}}{\lambda_a^{(l)}(\lambda^{(l)}_a+\lambda^{(l)}_b)}+\quad\quad\nonumber\\
 	&&+\sum_{a,b=1}^{N}\frac{(W^{(l)})_{ii}^{-1}.\Big(T^{(0)}_{<a,b>}\Big)_{l+i,l+i}}{\lambda_a^{(0)}(\lambda^{(0)}_a+\lambda^{(0)}_b)}+
 	\sum_{a,b=1}^{N}\sum_{c,d=1}^{N-l}\frac{\Big(T^{(0)}_{<a,b>}\Big)_{l+i,l+i}.\Big(T^{(l)}_{<c,d>}\Big)_{ii}}{\lambda_{a}^{(0)}\lambda_{c}^{(l)}(\lambda_{a}^{(0)}+\lambda_{b}^{(0)}+\lambda_{c}^{(l)}+\lambda_{d}^{(l)})}\Big].\nonumber\\
 	&&\int_{-\infty}^{0}d\tau \Big(\mathbb{G}_{ii}^{(m)}\mathbb{G}_{ii}^{(l)}-G_{ii}^{(m)}G_{ii}^{(l)}\Big)=\frac{1}{4}\Big[\sum_{a,b=1}^{N-l}\frac{(W^{(m)})_{ii}^{-1}.\Big(T^{(l)}_{<a,b>}\Big)_{ii}}{\lambda_a^{(l)}(\lambda^{(l)}_a+\lambda^{(l)}_b)}+\quad\quad\nonumber\\
 	&&+\sum_{a,b=1}^{N-m}\frac{(W^{(l)})_{ii}^{-1}.\Big(T^{(m)}_{<a,b>}\Big)_{ii}}{\lambda_a^{(m)}(\lambda^{(m)}_a+\lambda^{(m)}_b)}
 	+\sum_{a,b=1}^{N-m}\sum_{c,d=1}^{N-l}\frac{\Big(T^{(m)}_{<a,b>}\Big)_{ii}.\Big(T^{(l)}_{<c,d>}\Big)_{ii}}{\lambda_{a}^{(m)}\lambda_{c}^{(l)}(\lambda_{a}^{(m)}+\lambda_{b}^{(m)}+\lambda_{c}^{(l)}+\lambda_{d}^{(l)})}\Big]\nonumber\\
 	&&\int_{-\infty}^{0}d\tau \Big(\mathbb{G}_{i-l,i-l}^{(l)}\mathbb{G}_{i-m,i-m}^{(m)}-G_{i-l,i-l}^{(l)}G_{i-m,i-m}^{(m)}\Big)=
 \frac{1}{4}\Big[\sum_{a,b=1}^{N-l}\frac{\Big(W^{(m)}\Big)^{-1}_{i-m,i-m} .\Big(T^{(l)}_{<a,b>}\Big)_{i-l,i-l}}{\lambda_a^{(l)}(\lambda^{(l)}_a+\lambda^{(l)}_b)}\nonumber\\
 	&&+\sum_{a,b=1}^{N-m}\frac{(W^{(l)})_{i-l,i-l}^{-1}.\Big(T^{(m)}_{<a,b>}\Big)_{i-m,i-m}}{\lambda_a^{(m)}(\lambda^{(m)}_a+\lambda^{(m)}_b)}
 +\sum_{a,b=1}^{N-m}\sum_{c,d=1}^{N-l}\frac{\Big(T^{(m)}_{<a,b>}\Big)_{i-m,i-m}.\Big(T^{(l)}_{<c,d>}\Big)_{i-l,i-l}}{\lambda_{a}^{(m)}\lambda_{c}^{(l)}(\lambda_{a}^{(m)}+\lambda_{b}^{(m)}+\lambda_{c}^{(l)}+\lambda_{d}^{(l)})}\Big].\nonumber
 \end{eqnarray}
}
 	{\small
 		\begin{eqnarray}
 	&&\int_{-\infty}^{0}d\tau\Big( \mathbb{G}_{i,i-l}^{(m)}\mathbb{G}_{i-l,i-l+m}^{(l)}-G_{i,i-l}^{(m)}G_{i-l,i-l+m}\Big)=
 \frac{1}{4}\Big[\sum_{a,b=1}^{N-m}\frac{(W^{(l)})_{i-l,i-l+m}^{-1}.\Big(T^{(m)}_{<a,b>}\Big)_{i,i-l}}{\lambda_a^{(m)}(\lambda^{(m)}_a+\lambda^{(m)}_b)}+\quad\quad\nonumber\\
 	&&+\sum_{a,b=1}^{N-l}\frac{(W^{(m)})_{i,i-l}^{-1}.\Big(T^{(l)}_{<a,b>}\Big)_{i-l,i-l+m}}{\lambda_a^{(l)}(\lambda^{(l)}_a+\lambda^{(l)}_b)}
 	+\sum_{a,b=1}^{N-m}\sum_{c,d=1}^{N-l}\frac{\Big( T^{(m)}_{<a,b>}\Big)_{i,i-l}.\Big(T^{(l)}_{<c,d>}\Big)_{i-l,i-l+m}}{\lambda_{a}^{(m)}\lambda_{c}^{(l)}(\lambda_{a}^{(m)}+\lambda_{b}^{(m)}+\lambda_{c}^{(l)}+\lambda_{d}^{(l)})}\Big].\nonumber
\end{eqnarray}
}

 The above formulas, which are valid for the first-order interaction correction in the fuzzy sphere case, also apply to the fuzzy disc, with the exception that some of the sums are truncated and certain terms vanish. This follows from the fact that \(M^{(m)}\) vanishes for \(m>q-1\), since \(\mathbb{W}^{(m)}_- - \mathbb{W}^{(m)}_+\) vanishes for \(m>q-1\). The final expression for the first-order correction in the fuzzy disc is therefore given by

 {\small
 	\begin{eqnarray}
 	&&	\delta \ln \mathrm{Tr}\r_A^n=-\frac{n\lambda}{4!} \Bigg( 6\sum_{i=1}^{N}\Big[2\sum_{a,b=1}^{N}\frac{(W^{(0)})_{ii}^{-1} .\Big(T^{(0)}_{<a,b>}\Big)_{ii}}{\lambda_a^{(0)}(\lambda^{(0)}_a+\lambda^{(0)}_b)}+\sum_{a,b=1}^{N}\sum_{c,d=1}^{N}\frac{\Big(T^{(0)}_{<a,b>}\Big)_{ii}.\Big(T^{(0)}_{<c,d>}\Big)_{ii}}{\lambda_{a}^{(0)}\lambda_{c}^{(0)}(\lambda_{a}^{(0)}+\lambda_{b}^{(0)}+\lambda_{c}^{(0)}+\lambda_{d}^{(0)})}\Big]\nonumber\\
 		&+&4\sum_{l=1}^{q-1}\sum_{i=1}^{N-l}\Big[\sum_{a,b=1}^{N-l}\frac{(W^{(0)})_{l+i,l+i}^{-1}.\Big(T^{(l)}_{<a,b>}\Big)_{ii}}{\lambda_a^{(l)}(\lambda^{(l)}_a+\lambda^{(l)}_b)}+
 		\sum_{a,b=1}^{N}\sum_{c,d=1}^{N-l}\frac{\Big(T^{(0)}_{<a,b>}\Big)_{l+i,l+i}.\Big(T^{(l)}_{<c,d>}\Big)_{ii}}{\lambda_{a}^{(0)}\lambda_{c}^{(l)}(\lambda_{a}^{(0)}+\lambda_{b}^{(0)}+\lambda_{c}^{(l)}+\lambda_{d}^{(l)})}\Big]\nonumber\\
 		&+&4\sum_{l=1}^{N-1}\sum_{i=1}^{N-l}\sum_{a,b=1}^{N}\frac{(W^{(l)})_{ii}^{-1}.\Big(T^{(0)}_{<a,b>}\Big)_{l+i,l+i}}{\lambda_a^{(0)}(\lambda^{(0)}_a+\lambda^{(0)}_b)}+
 		8\sum_{m=1}^{q-1}\sum_{l=0}^{N-m-1}\sum_{i=1}^{N-l-m}\sum_{a,b=1}^{N-m}\frac{(W^{(l)})_{ii}^{-1}.\Big(T^{(m)}_{<a,b>}\Big)_{l+i,l+i}}{\lambda_a^{(m)}(\lambda^{(m)}_a+\lambda^{(m)}_b)}\nonumber\\
 		&+&8\sum_{m=1}^{N-1}\sum_{l=0}^{min(q-1,N-m-1)}\sum_{i=1}^{N-m-l}\sum_{a,b=1}^{N-l}\frac{(W^{(m)})_{l+i,l+i}^{-1}.\Big(T^{(l)}_{<a,b>}\Big)_{ii}}{\lambda_a^{(l)}(\lambda^{(l)}_a+\lambda^{(l)}_b)}\nonumber\\
 		&+&8\sum_{m=1}^{q-1}\sum_{l=0}^{min(q-1,N-m-1)}\sum_{i=1}^{N-m-l}\sum_{a,b=1}^{N-m}\sum_{c,d=1}^{N-l}\frac{\Big(T^{(m)}_{<a,b>}\Big)_{l+i,l+i}\Big(T^{(l)}_{<c,d>}\Big)_{ii}}{\lambda_{a}^{(m)}\lambda_{c}^{(l)}(\lambda_{a}^{(m)}+\lambda_{b}^{(m)}+\lambda_{c}^{(l)}+\lambda_{d}^{(l)})}\Big]\nonumber\\
 		&+&4\sum_{m=1}^{q-1}\sum_{l=1}^{m}\sum_{i=1}^{N-m}\sum_{a,b=1}^{N-m}\frac{(W^{(l)})_{ii}^{-1}.\Big(T^{(m)}_{<a,b>}\Big)_{ii}}{\lambda_a^{(m)}(\lambda^{(m)}_a+\lambda^{(m)}_b)}
 		+4\sum_{m=1}^{N-1}\sum_{l=1}^{min(m,q-1)}\sum_{i=1}^{N-m}\sum_{a,b=1}^{N-l}\frac{(W^{(m)})_{ii}^{-1}.\Big(T^{(l)}_{<a,b>}\Big)_{ii}}{\lambda_a^{(l)}(\lambda^{(l)}_a+\lambda^{(l)}_b)}+\nonumber\\
 		&+&4\sum_{m=1}^{q-1}\sum_{l=1}^{min(m,q-1)}\sum_{i=1}^{N-m}\sum_{a,b=1}^{N-m}\sum_{c,d=1}^{N-l}\frac{\Big(T^{(m)}_{<a,b>}\Big)_{ii}\Big(T^{(l)}_{<c,d>}\Big)_{ii}}{\lambda_{a}^{(m)}\lambda_{c}^{(l)}(\lambda_{a}^{(m)}+\lambda_{b}^{(m)}+\lambda_{c}^{(l)}+\lambda_{d}^{(l)})}\nonumber\\
 		&+&4\sum_{m=1}^{q-2}\sum_{l=m+1}^{q-1}\sum_{i=1}^{N-l}\Big[\sum_{a,b=1}^{N-l}\frac{(W^{(m)})_{ii}^{-1}.\Big(T^{(l)}_{<a,b>}\Big)_{ii}}{\lambda_a^{(l)}(\lambda^{(l)}_a+\lambda^{(l)}_b)}
 		+\sum_{a,b=1}^{N-m}\sum_{c,d=1}^{N-l}\frac{\Big(T^{(m)}_{<a,b>}\Big)_{ii}.\Big(T^{(l)}_{<c,d>}\Big)_{ii}}{\lambda_{a}^{(m)}\lambda_{c}^{(l)}(\lambda_{a}^{(m)}+\lambda_{b}^{(m)}+\lambda_{c}^{(l)}+\lambda_{d}^{(l)})}\Big]\nonumber\\
 		&+&4\sum_{m=1}^{q-1}\sum_{l=m+1}^{N-1}\sum_{i=1}^{N-l}\sum_{a,b=1}^{N-m}\frac{(W^{(l)})_{ii}^{-1}.\Big(T^{(m)}_{<a,b>}\Big)_{ii}}{\lambda_a^{(m)}(\lambda^{(m)}_a+\lambda^{(m)}_b)}
 		+4\sum_{m=1}^{q-1}\sum_{l=0}^{m}\sum_{i=m+1}^{N}\Big[\sum_{a,b=1}^{N-m}\frac{(W^{(l)})_{i-l,i-l}^{-1}.\Big(T^{(m)}_{<a,b>}\Big)_{i-m,i-m}}{\lambda_a^{(m)}(\lambda^{(m)}_a+\lambda^{(m)}_b)}\nonumber\\
 		&+&\sum_{a,b=1}^{N-m}\sum_{c,d=1}^{N-l}\frac{\Big(T^{(m)}_{<a,b>}\Big)_{i-m,i-m}.\Big(T^{(l)}_{<c,d>}\Big)_{i-l,i-l}}{\lambda_{a}^{(m)}\lambda_{c}^{(l)}(\lambda_{a}^{(m)}+\lambda_{b}^{(m)}+\lambda_{c}^{(l)}+\lambda_{d}^{(l)})}\Big]\nonumber\\
 		&+&4\sum_{m=1}^{N-1}\sum_{l=0}^{min(m,q-1)}\sum_{i=m+1}^{N}\sum_{a,b=1}^{N-l}\frac{(W^{(m)})_{i-m,i-m}^{-1}.\Big(T^{(l)}_{<a,b>}\Big)_{i-l,i-l}}{\lambda_a^{(l)}(\lambda^{(l)}_a+\lambda^{(l)}_b)}\nonumber\\
 &+&4\sum_{m=1}^{q-2}\sum_{l=m+1}^{q-1}\sum_{i=l+1}^{N}\Big[\sum_{a,b=1}^{N-l}\frac{(W^{(m)})_{i-m,i-m}^{-1}.\Big(T^{(l)}_{<a,b>}\Big)_{i-l,i-l}}{\lambda_a^{(l)}(\lambda^{(l)}_a+\lambda^{(l)}_b)}\nonumber\\
 		&+&\sum_{a,b=1}^{N-m}\sum_{c,d=1}^{N-l}\frac{\Big(T^{(m)}_{<a,b>}\Big)_{i-m,i-m}.\Big(T^{(l)}_{<c,d>}\Big)_{i-l,i-l}}{\lambda_{a}^{(m)}\lambda_{c}^{(l)}(\lambda_{a}^{(m)}+\lambda_{b}^{(m)}+\lambda_{c}^{(l)}+\lambda_{d}^{(l)})}\Big]
 		+4\sum_{m=1}^{q-1}\sum_{l=m+1}^{N-1}\sum_{i=l+1}^{N}\sum_{a,b=1}^{N-m}\frac{(W^{(l)})_{i-l,i-l}^{-1}.\Big(T^{(m)}_{<a,b>}\Big)_{i-m,i-m}}{\lambda_a^{(m)}(\lambda^{(m)}_a+\lambda^{(m)}_b)}\nonumber\\
 		&+&8\sum_{m=0}^{N-2}\sum_{l=1}^{min(q-1,N-m-1)}\sum_{i=l+1}^{N-m}\sum_{a,b=1}^{N-l}\frac{(W^{(m)})_{i,i-l}^{-1}.\Big(T^{(l)}_{<a,b>}\Big)_{i-l,i-l+m}}{\lambda_a^{(l)}(\lambda^{(l)}_a+\lambda^{(l)}_b)}+\nonumber\\
 		&+&8\sum_{m=0}^{q-1}\sum_{l=1}^{N-m-1}\sum_{i=l+1}^{N-m}\sum_{a,b=1}^{N-m}\frac{(W^{(l)})_{i-l,i-l+m}^{-1}.\Big(T^{(m)}_{<a,b>}\Big)_{i,i-l}}{\lambda_a^{(m)}(\lambda^{(m)}_a+\lambda^{(m)}_b)}+\nonumber
 			\end{eqnarray}
 	}
 {\small
\begin{eqnarray}
 		&+&	8\sum_{m=0}^{q-1}\sum_{l=1}^{min(q-1,N-m-1)}\sum_{i=l+1}^{N-m}\sum_{a,b=1}^{N-m}\sum_{c,d=1}^{N-l}\frac{\Big(T^{(m)}_{<a,b>}\Big)_{i,i-l}\Big(T^{(l)}_{<c,d>}\Big)_{i-l,i-l+m}}{\lambda_{a}^{(m)}\lambda_{c}^{(l)}(\lambda_{a}^{(m)}+\lambda_{b}^{(m)}+\lambda_{c}^{(l)}+\lambda_{d}^{(l)})} \Bigg).\nonumber
 	\end{eqnarray}
 }

 where $q$ denots the dimension of the subdisk; such that $R_s=\sqrt{q\theta}$ where $R_s$ is the radius of the subdisk.
 
 \section{ Construction of Zero-Mode-Projected Green's Function}\label{Appendix II}
In this section, we outline the main steps leading to the construction of the Green's function for the $(m=0)$ sector after projecting out the zero mode.

The Green's matrix function must satisfy the equation
\be\label{GFezmappendix}
\big[-\frac{d}{d\tau}M(\tau) \frac{d}{d\tau} +V(\tau) \big]\tilde{G}(\tau,\tau')=\delta (\tau-\tau') \mathbb{I}.
\ee

where
$$
M= \theta (-\tau) M_-+ \theta (\tau) M_+,~~
V(\tau)= \theta (-\tau) V_-+ \theta (\tau) V_+ .
$$

Here, $M_{\pm}$ and $V_{\pm}$ are arbitrary $N\times N$ positive-definite matrices.

By integrating \eqref{GFezmappendix} across the cut at $\tau=0$, it is straightforward to see that $\tilde{G}$ and its first derivative must satisfy the following continuity conditions:
$$
M_-\partial_\tau \tilde{G}_{-+}(0^-,\tau')=
M_+\partial_\tau \tilde{G}_{++}(0^+,\tau'),~~
\tilde{G}_{-+}(0^-,\tau')=
\tilde{G}_{++}(0^+,\tau'),~~ \tau' >0 .
$$

Similarly, for $\tau'<0$, we have
$$
M_+\partial_\tau \tilde{G}_{+-}(0^+,\tau')=
M_-\partial_\tau \tilde{G}_{--}(0^-,\tau'),~~
\tilde{G}_{--}(0^-,\tau')=
\tilde{G}_{+-}(0^+,\tau'),~~\tau'<0 .
$$

Consider first the $++$ region, for which \eqref{GFezmappendix} reduces to
$$
\big[-M_+\frac{d^2}{d\tau^2} +{V}_+ \big]\tilde{G}_{++}
=\delta(\tau-\tau') \mathbb{I}.
$$

Let $S_+$ be the square root of $M_+$ and define
$$
{G}= S_+ \tilde{G} S_+, ~~~
U_+= S_+^{-1} {V}_+ {S}_+^{-1} .
$$

It then follows that
$$
\big[-\frac{d^2}{d\tau^2} +U_+ \big]{{G}}_{++}
=\delta(\tau-\tau')\mathbb{I} .
$$

We are now essentially back to the same problem solved in \cite{Allouche_2022}. Using \eqref{GFS}, we can write down the general form of ${{G}}_{++}$, from which $\tilde{G}_{++}$ can be obtained. The remaining coefficients are then fixed by imposing the continuity conditions.

The solution in the $++$ region can be written as
$$
{{G}}_{++}= \frac{1}{2K_+} e^{-{K}_+|\tau-\tau'|}
+\frac{1}{2K_+} e^{-K_+\tau}
R_+ e^{-K_+ \tau'},~~~~
K_+=\sqrt{U_+}.
$$

Here, $R_+$ can be interpreted as a matrix-valued reflection coefficient, which is to be determined from the continuity conditions.

Similarly, the solution in the $--$ region is
$$
{{G}}_{--}= \frac{1}{2K_-} e^{-{K}_-|\tau-\tau'|}
+\frac{1}{2K_-} e^{K_-\tau}
R_- e^{K_- \tau'},~~~~
K_-=\sqrt{U_-}.
$$

The solutions in the mixed regions can be written as
$$
{{G}}_{\pm \mp}=
e^{\mp K_\mp\tau}
T_{\pm \mp}
e^{\pm K_\pm \tau'} .
$$

The matrices $T_{\pm\mp}$ and $R_-$ are also to be determined from the continuity conditions. They are not all independent, however, since $T_{+-}$ and $T_{-+}$ are related by Hermiticity.

Thus, we have\footnote{For $G_{+-}$ or $G_{-+}$, the differential equation requires only $G_{+-}= S_{+}^{-1}\tilde{G}_{+-} A$, with $A$ an arbitrary matrix. Hermiticity, however, requires $A=S_-^{-1}$, and similarly for $G_{-+}$.}
$$
\tilde{G}_{\pm \pm} =
S_{\pm}^{-1} {G}_{\pm \pm} S_{\pm}^{-1},~~~
\tilde{G}_{\pm \mp} =
S_{\pm}^{-1} {G}_{\pm \mp} S_{\mp}^{-1}.
$$

We now impose the continuity conditions. From
$$
G_{-+}(0^-,\tau')= G_{++}(0^+,\tau')
$$
we obtain
$$
S_+^{-1}\frac{1}{2K_+}(I+R_+)
=S_-^{-1}T_{-+}.
$$

Similarly,
$$
M_-\partial_\tau G_{-+}(0^-,\tau')
=M_+\partial_\tau G_{++}(0^+,\tau')
$$
gives
$$
S_+\frac{1}{2}(I-R_+)
=S_- K_-T_{-+}.
$$

Combining these two equations, we obtain
$$
T_{-+}=
(S_+^{-1}S_-K_-+K_+S_+S_-^{-1})^{-1},
$$

and
$$
\boxed{
	R_+=
	(K_+S_+S_-^{-1}-S_+^{-1}S_-K_-)
	(S_+^{-1}S_-K_-+K_+S_+S_-^{-1})^{-1}
}.
$$

Similarly, by interchanging $+$ and $-$, we obtain
$$
T_{+-}=
(S_-^{-1}S_+K_++K_-S_-S_+^{-1})^{-1},
$$

and
$$
\boxed{
	R_-=
	(K_-S_-S_+^{-1}-S_-^{-1}S_+K_+)
	(S_-^{-1}S_+K_++K_-S_-S_+^{-1})^{-1}
}.
$$

The expressions for the two sides are related simply by the interchange $+\leftrightarrow-$.

 \bibliographystyle{JHEP}
 \bibliography{document.bib}
\end{document}